\documentclass[12pt,a4paper]{article}
\usepackage{amsfonts}
\usepackage{amsmath}
\usepackage{bm}
\usepackage{fancyhdr}
\usepackage[a4paper, total={6.5in, 10in}]{geometry}
\usepackage{amssymb}
\usepackage{float}
\usepackage{caption}
\usepackage[caption=false]{subfig}
\usepackage{graphicx}
\usepackage{hyperref}
\usepackage{booktabs}
\usepackage{chngcntr}
\usepackage{placeins}
\usepackage{authblk}%
\usepackage{rotating}
\usepackage{tikz}
\usetikzlibrary{positioning}

\providecommand{\U}[1]{\protect\rule{.1in}{.1in}}

\newtheorem{theorem}{Theorem}

\newtheorem{proposition}[theorem]{Proposition}

\newenvironment{proof}[1][Proof]{\noindent\textbf{#1.} }{\ \rule{0.5em}{0.5em}}

\hypersetup{
  colorlinks   = true, 
  urlcolor     = blue, 
  linkcolor    = blue, 
  anchorcolor = blue
  citecolor   = blue 
}

\begin{document}

\title{FX-constrained growth: Fundamentalists, chartists and the dynamic trade-multiplier\thanks{An earlier version of this paper was presented at the 12th Dynamic Models in Economics and Finance Workshop, Urbino, Italy; and at the 13th Nonlinear Economic Dynamics Conference, Naples, Italy. We are grateful to the participants for their insightful comments and suggestions. Special thanks are due to L. Gardini, J. Jungeilges, I. Kubin,  G. Lima, T. Perevalova, G. Porcile, F. Westerhoff, and T. Zörner for the pleasant discussions while preparing an earlier draft of the article. The content is solely the responsibility of the authors, and usual caveats apply.}}
\author{Marwil J. Dávila-Fernández$^{\dagger \ddagger}$ \: $\cdot$ \: Serena Sordi$^{\ddagger}$}

\date{{\normalsize {$^{\dagger}$\emph{Colorado State University}}} \\
{\normalsize {$^{\ddagger}$\emph{University of Siena\smallskip}}\bigskip} \\
August 2025}


\maketitle

\begin{abstract}
Behavioural finance offers a valuable framework for examining foreign exchange (FX) market dynamics, including puzzles such as excess volatility and fat-tailed distributions. Yet, when it comes to their interaction with the `real' side of the economy, existing scholarship has overlooked a critical feature of developing countries. They cannot trade in their national currencies and need US dollars to access modern production techniques as well as maintain consumption patterns similar to those of wealthier societies. To address this gap, we present a novel heterogeneous agents model from the perspective of a developing economy that distinguishes between speculative and non-speculative sectors in the FX market. We demonstrate that as long as non-speculative demand responds to domestic economic activity, a market-clearing output growth rate exists that, in steady-state, is equal to the ratio between FX supply growth and the income elasticity of demand for foreign assets, i.e., a generalised dynamic trade-multiplier. Numerical simulations reproduce key stylised facts of exchange rate dynamics and economic growth, including distributions that deviate from the typical bell-shaped curve. Data from a sample of Latin American countries reveal that FX fluctuations exhibit similar statistical properties. Furthermore, we employ time-varying parameter estimation techniques to show that the dynamic trade-multiplier closely tracks observed growth rates in these economies.

\bigskip

 \bigskip\ 


\end{abstract}

\newpage

\section{Introduction}

A coherent description of foreign exchange (FX) market dynamics and their complex interplay with macroeconomic variables has long posed a significant challenge for the economics profession. Despite the predominance of the rational expectations efficient market model (e.g. \href{#Dornbusch 1976}{Dornbusch, 1976}; \href{#Valchev 2020}{Valchev, 2020}; \href{#Candian and De Leo 2023}{Candian and De Leo, 2023}), current assessments of the mechanisms driving FX movements and the interaction with the `real' side of the economy leave the picture somehow incomplete. Critical empirical puzzles, such as their pronounced volatility and fat-tailed distributions, highlight the need to continue looking for innovative approaches that integrate micro and macroeconomic dimensions.

Motivated by these exchange rate regularities, the literature on behavioural finance has provided an alternative to the rational expectations approach that highlights the role of boundedly rational heterogeneous agents who use simple decision heuristics (see \href{#De Grauwe and Grimaldi 2006}{De Grauwe and Grimaldi, 2006}; \href{#De Grauwe and Kaltwasser 2012}{De Grauwe and Kaltwasser, 2012}; \href{#Dieci and Westerhoff 2013}{Dieci and Westerhoff, 2013}; \href{#Gardini et al 2022}{Gardini et al., 2022}; \href{#Mignot and Westerhoff 2025}{Mignot and Westerhoff, 2025}). The recognition that the coexistence of different beliefs might introduce non-linear features into exchange rates dates back to \href{#Frankel and Froot 1986}{Frankel and Froot (1986)} and has been further developed primarily in the context of stock prices (for a review of the literature, see \href{#Dieci and He 2018}{Dieci and He, 2018}). A limited but growing number of studies have focused on the macro dynamic implications of heterogeneity in FX trading rules. Some of them have adopted a more business cycle set-up (e.g. \href{#Proano 2011}{Proaño, 2011}; \href{#Flaschel et al 2015}{Flaschel et al., 2015}; \href{#Gori and Ricchiuti 2018}{Gori and Ricchiuti, 2018}; \href{#Jang 2020}{Jang, 2020}) while others include growth-cycle considerations (as in \href{#Belloc and Federici 2010}{Belloc and Federici, 2010}; \href{#Delli Gatti et al 2024}{Delli Gatti et al., 2024}).

However, it should be noted that existing behavioural scholarship on the topic does not properly distinguish between developed and developing countries. If something, they have been treated as similar systems in different positions of a linear development path. This poses a problem, as countries like Argentina, Brazil, or Mexico cannot trade in their domestic currency. These economies occupy a peripheral position in the international monetary system and lag in technological capabilities, a situation that has been aggravated by the fact that the United States dollar (USD) has become increasingly central as the de facto reference currency for much of the world (\href{#Ilzetzki et al 2022}{Ilzetzki et al., 2022}). The bottom line is that most of the Global South needs internationally accepted currency to access modern production techniques and proxy consumption patterns of richer societies.\footnote{This feature is closely related to the inability to borrow abroad in national currency, i.e., the \textit{original sin}. Initiatives such as the Russian SPFS and the BRICS-PAY are attempts to bypass this constraint. Still, they are in relatively early stages of maturation and typically appear as a complement to USD operations. Despite some evidence of a decline in the dollar’s share of allocated foreign exchange reserves, reports of its demise as the dominant global currency remain, for the moment, highly exaggerated (Eichengreen, 2023).}

The present paper innovates by presenting a novel heterogeneous agent model designed from the perspective of a developing country. Its financial part dwells on \href{#Mignot and Westerhoff 2025}{Mignot and Westerhoff (2025)} revisitation of the original \href{#De Grauwe and Grimaldi 2006}{De Grauwe and Grimaldi (2006)} model, but comes with a critical difference: we distinguish between speculative and non-speculative sectors in the FX market. Speculators are categorised as fundamentalists, chartists, or trend extrapolators, with their composition assumed to be fixed. However, the critical parameters that represent the distribution of trading strategies vary in our numerical experiments. To the best of our knowledge, we are the first to show that as long as non-speculative demand responds to domestic economic activity, a market-clearing output growth rate exists that, in steady-state, is equal to the ratio between the non-speculative FX supply growth rate and the income elasticity of demand for foreign assets, i.e., a generalised dynamic trade-multiplier. We propose a mechanism grounded in a disequilibrium modelling approach (\href{#Belloc and Federici 2010}{Belloc and Federici, 2010}) that explains why such a growth rate serves as a reference point for the economy.

Numerical simulations indicate that the model is compatible with exchange rate and growth-stylised facts, such as high volatility and series that do not follow a normal distribution. This happens both for the stochastic version of the system without endogenous fluctuations and for the deterministic skeleton of the map after a Flip bifurcation occurred. The model admits a unique equilibrium only when there is no chartism as a trading strategy. Otherwise, there are two additional equilibria, one with the exchange rate undervalued and the other overvalued. In both cases, the output growth rate is equal to the dynamic trade multiplier. Studying the basins of attraction reveals the presence of multiple disconnected regions such that a strong exogenous exchange rate depreciation can paradoxically lead the system to the attracting region of the overvalued exchange rate. Thus, the model also provides an alternative explanation for the experience of Latin American (LA) countries that, after substantial depreciations in the late 1990s, have documented relatively overvalued currencies (\href{#Rodrik 2008}{Rodrik, 2008}; \href{#Ugurlu and Razmi 2023}{Ugurlu and Razmi, 2023}; for literature reviews, see \href{#Rapetti 2020}{Rapetti, 2020}; \href{#Demir and Razmi 2021}{Demir and Razmi, 2021}).

The obtained aperiodic fluctuations have the following rationale.
Suppose that the FX is depreciated relative to the fundamental. As long as the deviation is small, chartists' demand for more foreign currency leads to a further devaluation of the exchange rate. Given that the speculative sector absorbs the available USDs, there are fewer funds for non-speculative purposes. The real sector must adjust investment plans, decelerating output growth. The poor performance of the `real' economy is interpreted as a deterioration of economic fundamentals. Thus, agents expect a more depreciated FX fundamental, consolidating the initial deviation but without closing the gap with respect to the actual exchange rate. Trend-extrapolators attenuate volatility but lead to more persistent deviations. When they become too large, fundamentalists' response dominates, and the demand for USDs by the speculative sector falls strongly, leading to an overvaluation of the domestic currency. At that point, more resources become available to the non-speculative sector. Firms increase their capital accumulation, raising output growth. There is a perception that the economic fundamentals are improving, resulting in an appreciation of the expected fundamental rate, which now consolidates the previous appreciation of the FX, though not enough to close the gap. Chartists deepen the exchange rate overvaluation until it becomes too wide, triggering fundamentalists to buy in massive quantities. An exchange rate depreciation thus restarts the process.

In addition to the literature on behavioral finance, our framework also dialogues with two additional families of models. First, there is a clear-cut parallel between the proposed market-clearing output growth rate and the dynamic trade-multiplier (\href{#McCombie and Thirlwall 1997}{McCombie and Thirlwall, 1997}; for a literature review, see \href{#Blecker 2022}{Blecker, 2022}). The latter is usually obtained as the rate of growth that brings equilibrium to the current account, being equal to the growth rate of exports over the income elasticity of imports. Originally conceived to explain uneven development in the periphery of the economic system (see \href{#Prebisch 1959}{Prebisch, 1959}; \href{#Thirlwall and Hussain 1982}{Thirlwall and Hussain, 1982}), the trade multiplier has received significant single and multi-country empirical validation (e.g. \href{#Bagnai et al 2016}{Bagnai et al., 2016}; \href{#Kvedaras et al 2020}{Kvedaras et al., 2020}; \href{#Felipe and Lanzafame 2020}{Felipe and Lanzafame, 2020}; \href{#Srdelic and Davila-Fernandez 2025}{Srdelic and Davila-Fernandez, 2025}). We provide a novel generalisation that shows it can be obtained as a steady-state solution in the foreign exchange market. Furthermore, we are the first to use time-varying parameter estimation techniques to show that the dynamic trade-multiplier closely tracks observed growth rates in LA.

Second, the narrative developed here shares some elements with larger-scale agent-based models (see \href{#Dosi et al 2019}{Dosi et al., 2019}; \href{#Bassi et al 2023}{Basi et al., 2023}; \href{#Delli Gatti et al 2024}{Delli Gatti et al., 2024}). We are particularly close to Delli Gatti and coauthors, who differentiate between chartism and fundamentalism behaviour following \href{#De Grauwe and Grimaldi 2006}{De Grauwe and Grimaldi (2006)}. Despite their acknowledgement of some inspiration from the trade multiplier literature, they maintain the standard assumption that the exchange rate is the market-clearing critical variable. We argue this might not be the case in developing economies, or at least there is an additional ingredient that requires further consideration: output growth adjustments. Our tractable low-dimensional dynamic model complements those studies and is a first step towards a more extensive setup with a higher level of granularity.

The remainder of the paper is organised as follows. Section 2 overviews some exchange rate stylised facts and evidence supporting the empirical relevance of the trade-multiplier for a sample of Latin American countries. Section 3 presents our FX behavioural model, which consists of four main blocks of equations: foreign asset demand, supply, market clearing, and production technology. The resulting 2-dimensional nonlinear map is studied analytically and numerically in Section 4. We report additional robustness checks by varying the number of trading strategies. Numerically, it is shown that our main results are qualitatively preserved. We conclude with some final considerations.

\section{Exchange rates and the trade-multiplier}

\subsection{A quick look at FX dynamics in Latin America}

\begin{sidewaysfigure}[p]
\centering
  \includegraphics[width=9in]{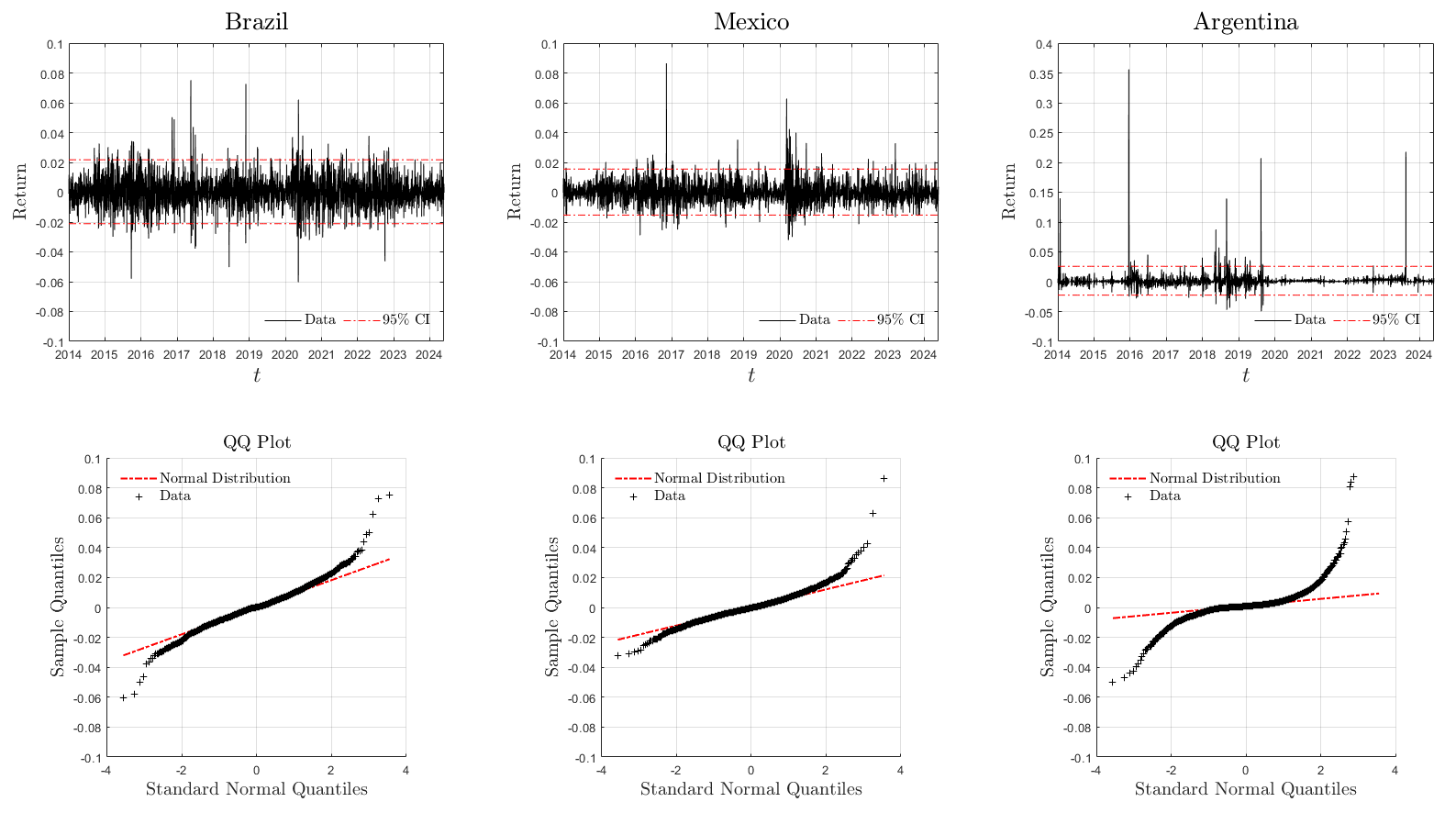} \bigskip
  \caption{High volatility and fat tails in exchange rates. Daily data for Brazil, Mexico, and Argentina, 2014-2024. Argentina series omits the max-depreciation of +120\% on December 14, 2023, which would make skewness and kurtosis uninformative higher. Anderson-Darling test indicates series do not follow a normal distribution.}\label{fig1}
\end{sidewaysfigure}

We begin by turning our attention to the behaviour of three major exchange rates ($E$) in Latin America: Brazil, Mexico and Argentina, the region's largest economies. We express $E$ as the value of domestic currency against the USD. Fig. \ref{fig1} uses daily data from 2014-2024, collected from \href{https://finance.yahoo.com/currencies/}{Yahoo Finance}. The upper row shows, in black, the returns calculated as $E_t/E_{t-1}-1$. The red dotted line marks 95\% confidence intervals. We regard such exchange rate fluctuations as excessive. While the case of Argentina might not be surprising as the country has been fighting since the late 2000s to stabilise its currency, Brazil and Mexico certainly have stronger fundamentals and are much more stable from a macroeconomic point of view. The bottom row reports the respective QQ plots. The Anderson-Darling test indicates that the series do not follow a normal distribution. A similar pattern is observed if we turn to the \emph{Andean Three}: Colombia, Chile, and Peru, as indicated in the Empirical Appendix \ref{EmpiricAppendix}. This second group is relatively more open to international trade and capital flows due to their smaller size and, to some extent, the absence of manufacturing industries.
The conclusion again is that FX returns are quite volatile and are not normally distributed.

\subsection{Introducing the trade-multiplier}

While the foreign exchange market is characterised by trading volumes predominantly driven by speculative orders (\href{#Menkhoff and Taylor 2007}{Menkhoff and Taylor, 2007}; \href{#Chaboud et al 2023}{Chaboud et al., 2023}), those dynamics are not isolated from the `real' sector. FX markets are particularly critical in LA because the region occupies a peripherical position in two aspects. First, from a technological perspective, modern production techniques are not created in Mexico or Peru but rather in the United States, Europe, or, more recently, China. Second, from a financial perspective, their domestic currencies cannot be used to engage in international trade, mainly because they are not accepted internationally as a store of value. Therefore, they need USDs to access and use cutting-edge technologies.

Take the academic sphere as an illustration. Major publishers set their prices in USD and Euros. For instance, most computers run with a Windows operating system, and Microsoft Office is the standard word processor and spreadsheet program. Our Peruvian colleagues, for example, cannot pay for their journal subscriptions nor keep their computers running in Nuevos Soles ($S/$). Moreover, as the most important international conferences take place in the hubs of developed countries, researchers need an internationally accepted currency to participate and effectively take part in scientific networks. A similar logic applies to the operationality of firms in these economies. To have access to capital and to keep it running, they need a constant flow of foreign currency. Without it, the economy would just stop.\footnote{The consolidation of international copyright and patent systems has amplified the role of the US dollar in developing countries. Previously, the primary challenge was access to capital itself; today, even when capital is available, domestic firms must pay for the right to use increasingly critical intangible assets. The rise of intangible capital, e.g. intellectual property, software, and branding, has deepened dependence on dollar-denominated transactions and institutions.} In this context, the dynamic trade-multiplier offers a simple tractable framework to study the growth implications of this uneven centre-periphery structure, particularly useful from a developing country point of view (\href{#McCombir and Thirlwall 1997}{McCombie and Thirlwall 1997}; see also \href{#Blecker 2022}{Blecker, 2022}).

Let us abstract from a moment of speculative motives in FX markets. Suppose the supply of foreign assets, determined by exports, grows at the exogenous rate $\Delta z$. On the other hand, demand is determined by an aggregate non-speculative sector, i.e., imports, which is a function only of Gross Domestic Product (GDP). If, in equilibrium, supply is equal to demand, then their growth rates are also equal. Thus, it follows that the GDP rate of growth that makes such an equality possible ($\Delta y^{BP}$) is defined and given by:

\begin{equation*}
    \Delta y^{BP}=\frac{\Delta z}{\pi}
\end{equation*}
where $\pi>0$ is the income elasticity of imports. In this literature, $\Delta y_t^{BP}$ is referred to as the dynamic trade-multiplier or the balance-of-payments constraint growth rate (for a detailed derivation, see \href{#Thirlwall and Hussain 1982}{Thirlwall and Hussain, 1982}).

\begin{sidewaysfigure}[p]
\centering
  \includegraphics[width=9in]{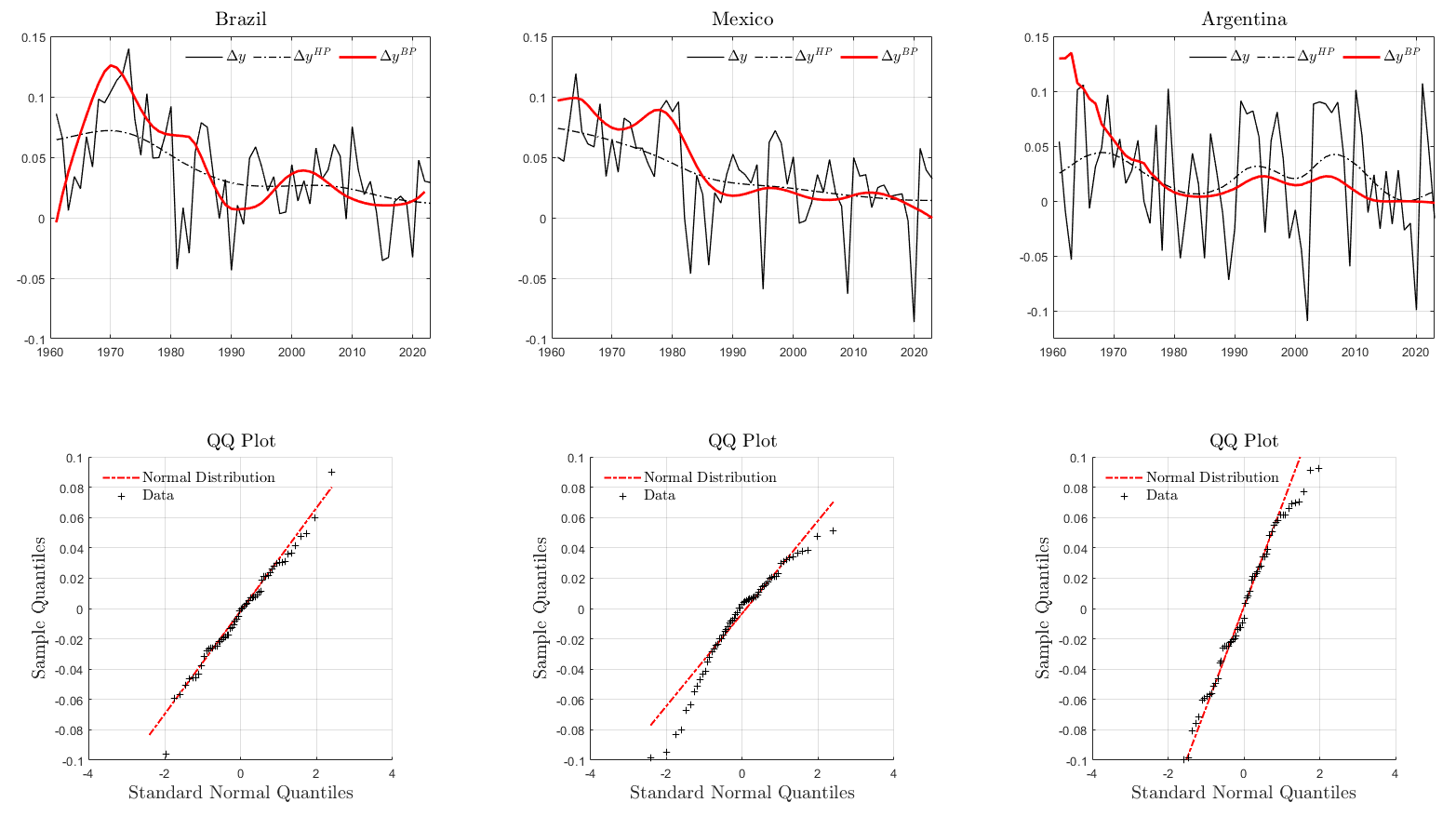} \bigskip
  \caption{Comparing actual growth rates ($\Delta y$) with the dynamic trade multiplier ($\Delta y^{BP}$). QQ plots indicate the difference $\Delta y - \Delta y^{BP}$ exhibits fat tails. Annual data for Brazil, Mexico, and Argentina, 1960-2023. Anderson-Darling test indicates series do not follow a normal distribution.}\label{fig3}
\end{sidewaysfigure}

The first row in Fig. \ref{fig3} reports, in red, our estimates of this simple rule for the three largest LA economies. We discuss our estimation procedure and report similar findings for Colombia, Chile and Peru in the Empirical Appendix \ref{EmpiricAppendix2}. The most important feature is that our red estimates closely follow the dotted black line, which marks the output growth trend ($\Delta y^{HP}$) obtained using the Hodrick–Prescott (HP) filter. Moreover, actual growth rates ($\Delta y$), indicated by the continuous black line, seem to fluctuate around the trade-multiplier. The latter exhibits a sort of negative trend separating before and after the 1980s, in line with studies documenting economic deceleration after the LA debt crisis (e.g. \href{#Cimoli et al 2010}{Cimoli et al., 2010}; \href{#Palma 2012}{Palma, 2012}). The second row in these figures calculates the QQ plot of the difference between $\Delta y - \Delta y^{BP}$. The Anderson-Darling test indicates that the series does not follow a normal distribution. We interpret this evidence as confirming, for example, \href{#Fagiolo et al 2008}{Fagiolo et al. (2008)} and \href{#Ascari et al 2015}{Ascari et al. (2015)}, who document fat tails in OECD growth rates (for a numerical application referring to the trade-multiplier, see \href{#Sordi and Davila-Fernandez 2022}{Sordi and Davila-Fernandez, 2022}).

Of course, FX dynamics are strongly influenced by demand and supply in the speculative sector. Their interplay with the trade-multiplier is particularly relevant when turning points in financial capital movements carry significant real economic consequences. They echo recent historical episodes in developing countries such as the Russian-Asian financial crisis of the late 1990s and Argentina’s struggles with the IMF. Taken together, Figs. (\ref{fig1}) and (\ref{fig3}) suggest that, at least for countries in the periphery of the global system, exchange rate and output dynamics might be more related than we initially thought. The model presented in the next section aims to integrate the trade multiplier into the traditional behavioural finance chartist-fundamentalist framework. To the best of our knowledge, we are the first to propose such a step. The speculative and non-speculative sectors will be micro-founded, ensuring that exchange rates and output growth are simultaneously determined. In equilibrium, all markets clear, but being compatible with persistent irregular fluctuations, ours is a disequilibrium growth cycle model.

\section{The model}

We consider an economy with an FX and a goods market. The former has speculative and non-speculative sectors that demand and supply foreign currency. Domestic and foreign speculators might behave as fundamentalists, chartists or extrapolators. Firms in the goods market operate with AK technology, allowing us to abstract from labour market considerations. They may or may not be able to adjust their capital accumulation plans depending on the rigidity of their contractual obligations. The model admits two connecting bridges between markets. On the one hand, expectations of the FX fundamentals are influenced by past economic performance. On the other hand, the FX market clearing output growth rate is a function of the current exchange rate and serves as a reference point for long-run economic performance, to which firms adjust their investment plans.\footnote{In this project, we face the important issue of different time scales between financial and real variables. Very high-frequency oscillations typically characterise exchange rate markets, whereas macroeconomic variables adjust over much lower frequencies. Trading decisions are made by the hour, while those related to production can take several weeks. Here, we will address this problem only indirectly in three complementary ways. First, the speed of adjustment to disequilibrium in FX markets will be much higher than in the `real' economy. Second, only a small fraction of firms will be allowed to adjust their capital stock every period. Finally, the expectations operator, carrying a stochastic component, will be used to represent the uncertainty behind agents' perceptions of the economic fundamentals. We leave an explicit treatment of the time scale issue for future research.}

The real exchange rate ($RER$) is defined as the nominal rate ($E$) expressed in domestic currency against the USD multiplied by the ratio between  foreign ($P^{\ast}$) and domestic prices ($P$), indicated as:
\begin{equation*}
    RER_t=E_t \frac{P_t^{\ast}}{P_t}
\end{equation*}
It implies:
\begin{align}
rer_{t} &  =e_t + p^{\ast}_t - p_t \label{rer}
\end{align}
where, unless stated otherwise, for any generic upper-case variable $X$, we indicate its natural logarithm using the lower-case $x= \ln X$, and approximate its growth rate by applying the first-difference operator ($\Delta$), writing $X_t/X_{t-1}-1=\Delta x_t$.\footnote{Our notation follows common practice in the macroeconomic literature. Notice that from its definition, we have that $\Delta x_t = x_t - x_{t-1}= \ln X_t - \ln X_{t-1}  =\ln (X_t/X_{t-1})$. The latter can be rewritten as $\ln \left( 1+(X_t-X_{t-1})/X_{t-1} \right) \approx (X_t-X_{t-1})/X_{t-1} = X_t/X_{t-1}-1$.} For our purposes, assume $p^{\ast}_t - p_t=0$ so that real and nominal exchange rates are the same, and we avoid the need to provide a description of inflationary processes.

\subsection{Market demand}

Market demand for foreign assets ($D$) is divided between an aggregate non-speculative sector ($D^{NS}$) and speculative demand ($D^{S}$), so that:
\begin{equation}
D_{t}=D_{t}^{NS}+D_{t}^{S}\label{MarketDemand}%
\end{equation}

The non-speculative sector consists of commercial traders that need to import from abroad. Following \href{#Thirlwall and Hussain 1982}{Thirlwall and Hussain (1982)}, but abstracting from the price component, we represent it as:
\begin{equation*}
D_{t}^{NS}= Y_t^{\pi}
\end{equation*}
where $Y$ is GDP and, as before, $\pi$ represents the response of import requirements to economic activity.
The expression above can be approximated in growth rates as:
\begin{equation}
\Delta d_{t}^{NS}= \pi \Delta y_t  \label{NS}%
\end{equation}

Regarding speculative demand, suppose there are $N$ speculative traders with different beliefs about the behaviour of the exchange rate. They operate from the perspective of a domestic national who has, for example, Brazilian Reais (R\$) or Mexican Pesos (Mex\$) and demands USDs. We write:
\begin{equation*}
D_{t}^{S}=\underset{i=1}{\overset{N}{\sum}}  D_{t}^{i}
\end{equation*}
implying in growth rates that:
\begin{equation}
\Delta d_{t}^{S}=\underset{i=1}{\overset{N}{\sum}} \phi^i \Delta d_{t}^{i}\label{S}%
\end{equation}
where the share of each trader $\phi^i \in (0,1)$ is such that $\sum^N_{i=1} \phi^i = 1$.

For the moment, we assume there are two types of speculative behaviour: Fundamentalists ($F$) and chartists ($C$). They provide a fair approximation of trading strategies in line with survey evidence on professional market participants (e.g. \href{#Menkhoff and Taylor 2007}{Menkhoff and Taylor, 2007}; \href{#Neely and Weller 2013}{Neely and Weller, 2013}; \href{#Hassanniakalager et al 2021}{Hassanniakalager et al., 2021}). The variation of speculator $i$'s order placement ($\Delta d$) in a given period is:\footnote{For simplicity, we assume the probability of adopting a fundamentalist or chartist heuristic is constant and exogenously given. While we acknowledge that, in reality, this is not the case, such an abstraction allows us to provide the simplest possible integration between the behavioural finance framework and the trade multiplier. A well-known alternative used, for example, by De Grauwe and Grimaldi (2006), is to rely on discrete-choice theory (see Brock and Hommes, 1997). We leave this step for future research.}
\begin{equation}
\Delta d_{t}^{i}=\left\{
\begin{array}
[c]{c}%
\Delta d_{t}^{F_{i}} \text{ with probability }w^{F_{i}}\\ \\
\Delta d_{t}^{C_{i}} \text{ with probability }w^{C_{i}}%
\end{array}
\right.  \label{Speculators}%
\end{equation}

Fundamentalists assess that deviations in the exchange rate from its fundamental value ($f$) will be corrected shortly. Therefore, they buy foreign currency-denominated financial assets when $\mathbb{E} [f]>e$ while selling under $\mathbb{E} [f]<e$. Hence:
\begin{equation}
\Delta d_{t}^{F_{i}}=\mu\left(  \mathbb{E} [f_{t}]-e_{t-1}\right)^3
\label{Fundamentalists}%
\end{equation}
where $\mu>0$ is a reaction parameter and $\mathbb{E[\cdot]}$ is the expectations operator. Traders must form expectations about the fundamental because its `true' value is unobservable or is observed only after the performance of the economy is determined. While FX trading and production decisions occur daily, macroeconomic statistics that inform the fundamental value are typically released on a monthly or quarterly basis. Consequently, the fundamental exchange rate is subject to significant uncertainty, reflecting underlying economic conditions that are only partially known during trading.\footnote{This assumption is part of our toolkit to deal with the issue of different time scales. We rely on two additional features to accommodate this problem. For instance, in our numerical experiments, the speed of adjustment to disequilibrium in FX markets is assumed to be much higher than that of the ‘real’ economy. Moreover, only a small fraction of firms can adjust their capital stock every period. We thank I. Kubin for noting that, as in Dornbusch (1976), the dynamics we obtain are linked to the assumption that exchange rates and asset markets adjust fast relative to goods markets.} Finally, the cubic term in Eq. (\ref{Fundamentalists}) captures the insight of \href{#Day and Huang 1990}{Day and Huang (1990)}, and those afterwards, that the strength of the response of this type of agent increases with the magnitude of the deviation relatively to the fundamental. The higher the frustration in expectations, the stronger their response will be.

\begin{figure}[tbp]
    \centering
    \includegraphics[width=0.4\linewidth]{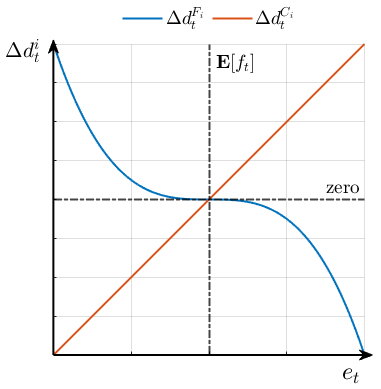}
    \caption{Domestic fundamentalists and chartists trading strategies.}
    \label{Heuristics}
\end{figure}

On the other hand, chartists anticipate some persistence in exchange rate deviations from the fundamental price. The literature provides at least two ways to formalise this strategy. First, we can directly write $\Delta d_{t}^{C_{i}} $ as a function of the difference between $e$ and $\mathbb{E} [f]$ (see, for example, \href{#Dieci and Westerhoff 2010}{Dieci and Westerhoff, 2010}; \href{#Gardini et al 2022}{Gardini et al., 2022}; a similar heuristic is referred to as `contrarians' in \href{#Gori and Ricchiuti 2018}{Gori and Ricchiuti, 2018}). Alternatively, one could assume they are trend-extrapolators whose demand adjusts to $\Delta e_{t-1}$ (e.g. \href{#De Grauwe and Grimaldi 2006}{De Grauwe and Grimaldi, 2006}; \href{#Proano 2011}{Proaño, 2011}; \href{#Federici and Gandolfo 2012}{Federici and Gandolfo, 2012}; \href{#Mignot and Westerhoff 2025}{Mignot and Westerhoff, 2025}). We will apply both in this presentation, beginning with the first formulation and leaving the second as a robustness check exercise. Hence, assume:

\begin{equation}
\Delta d_{t}^{C_{i}}=\mu\left(  e_{t-1}-\mathbb{E} [f_{t}]\right)
\label{Chartists}%
\end{equation}
so that their demand falls when $\mathbb{E} [f]>e$ and increases with $\mathbb{E} [f]<e$. Fig. \ref{Heuristics} depicts variation in fundamentalists' and chartists' demand as a function of deviations with respect to expectations.

If there is a continuum of speculators with mass $N$, we can express the market shares of speculators relying on fundamental or chartist analysis as the respective probabilities $w^{F_{i}}=w^{F}$ and $w^{C_{i}}=w^{C}$. To avoid excessive parameters, assume speculators are the same size so that $\phi^i=1/N$. From Eqs. (\ref{S}) and (\ref{Speculators}), it follows that their FX demand adjustment is such that:
\begin{equation}
\Delta d_{t}^{S}=  w_{t}^{F} \Delta d_{t}^{F}+w_{t}^{C} \Delta d_{t}^{C}
\label{SpeculatorsDemandPre}%
\end{equation}
Substituting the behavioural heuristics of fundamentalists and chartists in Eqs. (\ref{Fundamentalists}) and (\ref{Chartists}) into (\ref{SpeculatorsDemandPre}), then we have:%
\begin{equation}
\Delta d_{t}^{S}=\mu \left[ w^F \left(  \mathbb{E} [f_{t}]-e_{t-1}\right)^3+ w^C \left(e_{t-1}-\mathbb{E} [f_{t}]  \right)  \right] \label{S2}%
\end{equation}

\subsection{Market supply}

To describe the market supply of foreign assets ($Z$), we turn again to the coexistence of non-speculative ($Z^{NS}$) and speculative ($Z^{S}$) sectors:
\begin{equation*}
    Z_t=Z_t^{NS}+Z_{t}^{S}
\end{equation*}

Keeping the first of them as simple as possible, we suppose it grows at an exogenous rate $\Delta z^{NS}$. On the other hand, speculative supply follows a structure similar to and symmetric to demand. Notice, however, that agents now act from the perspective of a foreign national who owns USDs. Her/his decision is whether to exchange them for, e.g. Mex\$ or R\$. We write $Z^S$ as the sum of individual suppliers:
\begin{equation*}
Z_{t}^{S}=\underset{i=1}{\overset{N}{\sum}}  Z_{t}^{i}
\end{equation*}
implying in growth rates that:
\begin{equation}
\Delta z_{t}^{S}=\underset{i=1}{\overset{N}{\sum}} \phi^i \Delta z_{t}^{i}\label{S3}%
\end{equation}
where as before the share of each trader $\phi^i \in (0,1)$ is such that $\sum^N_{i=1} \phi^i = 1$.

There are $N$ traders that can be either fundamentalists or chartists. For simplicity, we assume probabilities are the same for those demanding and supplying in the FX market. Thus, the variation of speculator $i$'s order placement in a given period is:\begin{equation}
\Delta z_{t}^{i}=\left\{
\begin{array}
[c]{c}%
\Delta z_{t}^{F_{i}}\text{ with probability }w^{F_{i}}\\ \\
\Delta z_{t}^{C_{i}}\text{ with probability }w^{C_{i}}%
\end{array}
\right.  \label{Speculators2}%
\end{equation}

A foreign fundamentalist assesses that deviations in the exchange rate from expectations cannot persist over time. However, from her/his perspective, this means they sell USDs when $\mathbb{E} [f]<e$ and buy under $\mathbb{E} [f]>e$. Hence:
\begin{equation}
\Delta z_{t}^{F_{i}}=\rho\left( e_{t-1}- \mathbb{E} [f_{t}]\right)^3
\label{Fundamentalists2}%
\end{equation}
where $\rho>0$ is a reaction parameter and, for simplicity, will be taken as constant across agents and types of behaviour. As before, the cubic term captures their non-linear response to expectation deviations. 

Chartists anticipate some persistence in exchange rate misalignments. This strategy implies traders want to sell more USDs when $\mathbb{E} [f]>e$, buying if $\mathbb{E} [f]<e$. Therefore, assume:
\begin{equation}
\Delta z_{t}^{C_{i}}=\rho\left( \mathbb{E} [f_{t}] - e_{t-1}\right)
\label{Chartists2}%
\end{equation}

As before, suppose speculators are the same size so that $\phi^i=1/N$. From Eqs. (\ref{S3}) and (\ref{Speculators2}), it follows that their FX supply adjustment is such that:
\begin{equation}
\Delta z_{t}^{S}=  w^{F} \Delta z_{t}^{F}+w^{C} \Delta z_{t}^{C}
\label{SpeculatorsDemandPre2}%
\end{equation}
Making use of the behavioural heuristics of foreign fundamentalists and chartists in Eqs. (\ref{Fundamentalists2}) and (\ref{Chartists2}) into Eq. (\ref{SpeculatorsDemandPre2}), the expression above becomes:
\begin{equation}
\Delta z_{t}^{S}=\rho \left[ w^F \left( e_{t-1}- \mathbb{E} [f_{t}]\right)^3 + w^C \left(  \mathbb{E} [f_{t}]-e_{t-1}\right) \right]  \label{S4}%
\end{equation}

\subsection{Market clearing and the trade-multiplier}

In equilibrium, the market demand for foreign assets at time $t$ is the sum of sectoral demands equal to market supply:

\[
D_{t}=Z_{t}%
\]
which from Eq. (\ref{MarketDemand}) implies in growth rates:%
\begin{equation}
\theta \Delta d_{t}^{NS}+ \left( 1-\theta \right) \Delta d_{t}^{S}= \theta \Delta z^{NS} + \left( 1-\theta \right) \Delta z_{t}^{S}
\label{MarketEquilibrium}%
\end{equation}
where $\theta \in (0,1)$ is a parameter capturing the sectoral composition of the FX market.

Most of the behavioural finance literature dealing with real-financial interactions assumes the exchange rate is the variable that clears the market. For example, in \href{#Proano 2011}{Proaño (2011)} and \href{#Flaschel et al 2015}{Flaschel et al. (2015)}, FX adjusts to the interest rate differentials and risk expectations. Alternatively, in other specifications, $e$ brings the balance of payments to equilibrium (e.g. \href{#Belloc and Federici 2010}{Belloc and Federici, 2010}; \href{#Gori and Ricchiuti 2018}{Gori and Ricchiuti, 2018}). In the ABM developed by \href{#Delli Gatti et al 2024}{Delli Gatti et al. (2024)}, which has a structure that follows very closely \href{#De Grauwe and Grimaldi 2006}{De Grauwe and Grimaldi (2006)}, once more, the exchange rate is the adjustment element. However, in countries that cannot trade in their own currency, this might not be the case, or at least there is an additional ingredient to be considered.

The presence of a non-speculative sector whose demand for FX currency responds to domestic activity implies that there exists an output growth rate that clears the market ($ \Delta y^{MC}$). That is the paramount insight behind the trade multiplier. Substituting Eqs. (\ref{NS}), (\ref{S2}), (\ref{S4}) in Eq. (\ref{MarketEquilibrium}), and solving for $\Delta y$ at time $t$, we obtain $ \Delta y^{MC}$ defined and given by:
\begin{equation}
\Delta y_{t}^{MC}=\frac{\Delta z^{NS}}{\pi}-\left(  \frac{1-\theta}{\theta
}\right)  \left(  \frac{\mu+\rho}{\pi}\right)  \underset{\text{FX speculative
trade}}{\underbrace{\left[  w^{F}\left(  \mathbb{E}[f_{t}]-e_{t-1}\right)
^{3}+w^{C}\left(  e_{t-1}-\mathbb{E}[f_{t}]\right)  \right]  }}\label{yMC}%
\end{equation}

When exchange rate expectations are satisfied, $\mathbb{E}[f]=e$, there is no speculative trade and $\Delta y^{MC}$ equals the dynamic trade-multiplier ($\Delta y^{BP}$), that is:
\[
\Delta y^{BP}=\frac{\Delta z^{NS}}{\pi} 
\]
This is perhaps one of the central insights of the present paper. It shows that the trade multiplier can be obtained as a market-clearing condition in the exchange rate market instead of imposing equilibrium in the current account (as in \href{#McCombie and Thirlwall 1997}{McCombie and Thirlwall, 1997}; see also \href{#Blecker 2022}{Blecker, 2022}), provided that there is a non-speculative sector that responds to economic performance.

To obtain the dynamics of the exchange rate, we refer to the covered interest parity (CIP). It states that the interest rate differential between two countries is exactly offset by the forward premium in the foreign exchange market:
\begin{equation}
    r-r^*=f_t-e_t \label{CIP1}
\end{equation}
where $r$ and $r^*$ are the domestic and foreign interest rates, while $f$ is the forward exchange, i.e., the rate contracted now for delivery at a future time. To keep our narrative as simple as possible, we abstract from monetary policy considerations and assume interest rates are constant. Applying the first-difference operator on Eq. (\ref{CIP1}) and rearranging, it follows that:
\begin{equation}
   \Delta e_t= \Delta f_t \label{CIP2}
\end{equation}

We regard the forward exchange rate as primarily speculative, as non-financial corporations account for less than 10\% of FX forward turnover (\href{#BIS 2022}{BIS, 2022}). Thus, $\Delta f$ is assumed to depend on variations in speculative demand:
\begin{equation}
   \Delta f_t=\Delta d_{t}^{S} - \Delta z_{t}^{S}  \label{fe}
\end{equation}
Substituting Eqs. (\ref{S2}) and (\ref{S4}) into Eq. (\ref{fe}), and the resulting expression into (\ref{CIP2}), then: 
\begin{equation}
e_{t}=e_{t-1}+ (\mu + \rho) \left[ w^{F}\left(  \mathbb{E}[f_{t}]-e_{t-1}\right)
^{3}+w^{C}\left(  e_{t-1}-\mathbb{E}[f_{t}]\right)  \right]  \label{E2}%
\end{equation}

\subsection{Production technology}

To introduce endogenous growth without departing too much from the behavioural approach to exchange rates, we rely on \href{#Belloc and Federici 2010}{Belloc and Federici (2010)}. They developed a two-country model for the Euro-USD that formalises the economy's supply side using an $AK$ production technology. Suppose there are $M$ firms of different types $j$ such that the sum of individual productive units gives the total output:
\begin{equation}
Y_{t}=\overset{M}{\underset{j=1}{\sum}}A^{j}K_{t}^{j} \label{ak1}
\end{equation}
where $A$ is a technological coefficient that differs across firms, and $K$ is the capital stock. A more refined production technique, including labour market considerations, would increase the complexity of the model without adding much to our main narrative, which is primarily concerned with the dynamics of the exchange rate and the trade multiplier.

Log-differentiating Eq. (\ref{ak1}), we obtain:
\begin{equation}
\triangle y_{t}=\overset{M}{\underset{j=1}{\sum}}\phi^{j}\triangle k_{t}%
^{j}\label{deltay2}%
\end{equation}
where $\phi^{j}\in(0,1)$ is the share of each firm and $\sum_{j=1}^{M} \phi^{j}=1$; so aggregate growth results from the sum of all firm's capital accumulation efforts weighted by their size.

Returning to the observation that developing economies like Brazil, Mexico, or Argentina need international currency to access capital goods, an increase in output growth implies higher input requirements denominated in USD. If the US needs to import a machine from China, it can pay for it in its currency. However, if Brazil needs to do the same or purchase a software license from an American company, they cannot pay in R\$. Thus, keeping everything else constant, an acceleration in domestic economic activity will eventually result in an imbalance between the demand and supply of foreign assets. The economy might grow above $\Delta y^{MC}$ for some time, contingent upon the liquidity of financial markets. Still, the market clearing output growth becomes a reference point for the economy (the reader is referred again to the literature review by \href{#Blecker 2022}{Blecker, 2022}; a comparison between ‘binding constraint’ and ‘centre-of-gravity’ perspectives can be found in \href{#Davila-Fernandez and Sordi 2024}{Davila-Fernandez and Sordi, 2024}).

While persistent deviations from this benchmark are perceived as unsustainable, it is unlikely that all firms will be able to adjust simultaneously to $\Delta y^{MC}$. Thus, we assume two types of capital accumulation behaviour: flexible ($flex$) and rigid ($rig$). Every period $t$, firm $j$ adjusts its capital accumulation strategy according to:
\begin{equation}
\triangle k_{t}^{j}=\left\{
\begin{array}
[c]{c}%
\triangle k_{t}^{flex}\text{ with probability }w^{flex}\\ \\
\triangle k_{t}^{rig}\text{ with probability }w^{rig}%
\end{array}
\right.  \label{kprob}%
\end{equation}

Flexible firms can tune their capital stock to international liquidity conditions imposed by the difference between interest rates as well as $\triangle y^{MC}$ and $\triangle y$. When $\triangle y^{MC}> \triangle y$, there is an excess FX supply relative to demand that eases the access to capital and the means needed to keep it running. Alternatively, $\triangle y^{MC}< \triangle y$ stands for the opposite case, and firms revise investment plans downwards, reducing capital accumulation. Given that we are abstracting from the effects of monetary policy, their investment strategy can be written as:
\begin{equation}
\triangle k_{t}^{flex}= \triangle k_{t-1}^{flex} + \beta\left(  \triangle
y_{t}^{MC}-\triangle y_{t-1}\right)  \label{kflex}%
\end{equation}
where $0<\beta<1$ captures the productive unit's reaction speed.

Rigid firms would like to follow a similar strategy but cannot do so. There are at least three main reasons why this might be the case. First, existing contracts, regulatory approvals, and other legal obligations may bind firms to certain investment commitments, limiting their ability to adapt swiftly to changing economic conditions. Second, many investments, especially in physical capital, are irreversible or have high sunk costs, making firms cautious about changing plans abruptly. Adjustment costs include logistical challenges associated with halting or redirecting ongoing projects. Finally, firms may face uncertainty about whether changes in economic conditions are temporary or permanent. They may not have immediate access to reliable information, leading to delays in decision-making. The bottom line is that this group of firms keeps business as usual:
\begin{equation}
\triangle k_{t}^{rig}=\triangle k_{t-1}^{rig}\label{krig}%
\end{equation}
If there is a continuum of firms with mass $M$, assuming they are of the same size so that $\phi^{j}=1/M$, from Eqs. (\ref{deltay2}) and (\ref{kprob}), it follows that the output growth rate is such that:
\begin{equation*}
\triangle y_{t}=w^{flex}\triangle k_{t}^{flex}+w^{rig}\triangle k_{t}%
^{rig}\label{deltay3}%
\end{equation*}
Substituting Eqs. (\ref{kflex}) and (\ref{krig}) into the expression above, we have:
\begin{equation}
\triangle y_{t}=\triangle y_{t-1}+w^{flex}\beta\left(  \triangle y_{t}%
^{MC}-\triangle y_{t-1}\right)  \label{y}%
\end{equation}

\subsection{Expectations and the FX fundamental}

Lastly, we must determine how traders form their expectations about the fundamental exchange rate.\footnote{There is no clear-cut definition in the literature for the ``fundamental'' exchange rate. Different, often overlapping, means of explaining $f$ include references to ``the equilibrium exchange rate'' (De Grauwe and DeWachter, 1993, p. 356) or ``important macroeconomic fundamentals such as price levels and real income'' (Taylor, 1995, p. 30). Among behavioural approaches, Proaño (2011) uses purchasing power parity, Belloc and Federici (2010) and Federici and Gandolfo (2012) refer to a natural real exchange rate (NATREX), while Delli Gatti et al. (2024) takes it as the one that keeps trade balanced.} To keep the narrative as simple as possible, suppose agents use the PPP rate ($e_{PPP}$) as a reference point. Still, motivated by \href{#Taylor 1995}{Taylor (1995)}, we assume they also look at the real side of the economy. Following \href{#Westerhoff 2012}{Westerhoff's (2012)} insight that the fundamental price of an asset also reflects recent macroeconomic performance, we allow for an adjustment component in the expectations operator that accounts for past growth. In mathematical terms:
\begin{equation*}
\mathbb{E} [f_{t}]=e_{PPP}- \Omega \Delta y_{t-1} +\epsilon_t \label{f}%
\end{equation*}
where $0<\Omega <1$ is a scaling coefficient and
\begin{equation*}
    \epsilon_t \sim N(0,\sigma)
\end{equation*}
This specification is in line with our definition of $e$, which expresses the domestic currency in terms of USD. An economy that grows vigorously is perceived to have better fundamentals, resulting in a more appreciated rate. Stochastic shocks ($\epsilon$) are i.i.d., and $\sigma$ stands for standard errors. They capture the intrinsic uncertainty surrounding the formation of expectations, especially regarding the proper fundamentals of a developing country. Moreover, the `true' FX fundamental remains unobservable in real-time as macroeconomic statistics are typically released monthly or quarterly.

Recalling that we are abstracting from inflation considerations, i.e. $p_t-p^*_t=0$, it is easy to see from Eq. (\ref{rer}) that $e_{PPP}=0$. Thus, we can rewrite the expected fundamental as:
\begin{equation}
\mathbb{E} [f_{t}]= - \Omega \Delta y_{t-1} +\epsilon_t \label{f}%
\end{equation}

\section{Dynamic system}

Substituting Eq. (\ref{f}) into (\ref{E2}), traders respond to real variables in a relationship intermediated by growth. Moreover, by substituting Eq. (\ref{yMC}) into (\ref{y}), the output adjusts to disequilibrium in the exchange rate market. Abstracting from the stochastic component to focus on the deterministic skeleton of the model, we obtain a 2D first-order nonlinear map:
\begin{align}
e_{t} &  =e_{t-1}+ (\mu + \rho) \left[ w^{F}\left(  - \Omega \Delta y_{t-1}-e_{t-1}\right)
^{3}+w^{C}\left(  e_{t-1}+ \Omega \Delta y_{t-1}\right)  \right]
\nonumber\\
& \label{2D}\\
\triangle y_{t} &  =\triangle y_{t-1}+w^{flex}\beta \left\{ \triangle y^{BP} -\gamma  \left[  w^{F}\left(
- \Omega \Delta y_{t-1}-e_{t-1}\right)  ^{3}+w^{C}\left(  e_{t-1}+ \Omega \Delta y_{t-1} \right)  \right] -\triangle
y_{t-1}\right\}  \nonumber
\end{align}
where $\gamma= \left(  1-\theta\right) (\mu + \rho)
/\theta \pi > 0$.

In steady-state, $e_{t}=e_{t-1}=\bar{e}$ and
$\Delta y_{t}=\Delta y_{t-1}=\Delta \bar{y}$. The correspondent equilibrium conditions of the dynamic system (\ref{2D}) are given by:
\begin{align}
0 &  =w^{F}\left(  - \Omega \Delta \bar{y}-\bar{e}\right)
^{3}+w^{C}\left(  \bar{e}+ \Omega \Delta \bar{y}\right)  \nonumber\\
& \label{EqCond}\\
0 &  =\triangle y^{BP} -\gamma  \left[  w^{F}\left(
- \Omega \Delta \bar{y}-\bar{e}\right)  ^{3}+w^{C}\left(  \bar{e}+ \Omega \Delta \bar{y}\right)  \right] -\Delta \bar{y} \nonumber
\end{align}
\noindent
These expressions enable us to distinguish between three distinct sets of mechanisms. First, equilibrium in FX markets depends on the exchange rate being equal to the expected fundamental, subject to the key nonlinearity in one of the trading strategies. Second, equilibrium in the goods market is contingent upon a stable flow of USDs, which is achieved only when the actual growth rate equals the trade-multiplier. Finally, the expected fundamentals will ultimately be informed by $\triangle y^{BP}$.

\subsection{Existence and local stability of equilibria}

Thus, we can state and prove the following Propositions regarding the existence of a unique or multiple equilibrium solution.

\begin{proposition} \label{prop 1}
When there is no heterogeneity in FX speculative trading strategies, i.e. all agents are either fundamentalists or chartists, the dynamic system (\ref{2D}) has a unique market-clearing equilibrium point $P_1=(\bar{e}_1,\Delta \bar{y}_1)$ defined and given by:
\begin{align*}
\bar{e}_1  & = - \Omega \Delta \bar{y}\\
\Delta \bar{y}_1  & = \Delta y^{BP}%
\end{align*}
If $w^F=1 \wedge w^C=0$, then $P_1$ is locally stable. On the contrary, when $w^F=0 \wedge w^C=1$, the equilibrium solution is unstable.

\bigskip

\begin{proof}
See Appendix B.1.
\end{proof}

\end{proposition}

\medskip

\begin{proposition}
When there is heterogeneity in FX speculative trading strategies, i.e. $w^F \wedge w^C \neq 0$, the dynamic system (\ref{2D}) admits two additional solutions $P_2=(\bar{e}_2,\Delta \bar{y}_2)$ and $P_3=(\bar{e}_3,\Delta \bar{y}_3)$, such that:
\begin{align*}
\bar{e}_{2,3}  & = - \Omega \Delta \bar{y} \pm \sqrt{\frac{w^C}{w^F}} \qquad \qquad \qquad \Delta \bar{y}_{2,3}  = \Delta y^{BP}
\end{align*}
\label{prop 2}

\begin{proof}
See Appendix B.2.
\end{proof}

\end{proposition}

\bigskip

Notice that the deviation of the equilibrium exchange rate with respect to $e_{PPP}=0$ can be decomposed into two parts. One is $\Omega \bar{y}$. Higher growth is interpreted as stronger fundamentals, leading to a more appreciated FX. However, as $\Omega$ is likely quite small, this effect is relatively minor. The second, most important, lies in the presence of chartism behaviour. Not only are they responsible for the existence of the two additional equilibria, but the gap between $e_{2,3}$ and $e_1$ increases with the share $w^C$. Regarding the local stability of equilibria, we are now ready to prove the following Proposition.

\begin{proposition}
When there is heterogeneity in trading strategies, $P_1$ is a saddle point whereas $P_{2,3}$ are locally stable nodes in the region of the parameter space defined by:
\begin{align*}
    A=4-4\left(  \mu+\rho\right)  w^{C}+2 \left[ \left( \mu + \rho+ 2\gamma \Omega \right)w^C-1 \right]w^{flex} \beta>0
\end{align*}
and
\begin{align*}
B=\left(  1-2\gamma\Omega w^{C}\right)  w^{flex}\beta+2\left(  \mu
+\rho\right)  w^{C}\left(  1-w^{flex}\beta\right)>0
\end{align*}
If a change in one of the parameters results in a violation of only the first condition, a Flip bifurcation occurs. On the other hand, a violation of the second condition while the first holds is associated with a Neimark-Sacker bifurcation. 
\label{prop 3}

\bigskip

\begin{proof}
See Appendix B.3.
\end{proof}
\end{proposition}

Despite having only one nonlinearity, the model is compatible with a rich range of behaviours, including endogenous, persistent, and irregular fluctuations. Oscillations generated in the FX market affect productive economic decisions, which in turn inform expectations about economic fundamentals. The economic rationale of such movements will be explored in more detail through a set of numerical simulations.


\subsection{Numerical experiments}

\begin{table}
    \centering
    \caption{Choice of parameters} \label{parameters}
\begin{tabular}{ccc}
\hline\hline
{\scriptsize Parameter} & {\scriptsize Value} & {\scriptsize Reference} \\ 
\hline
\scriptsize $\mu$ & \scriptsize $0-15$ & {\scriptsize Mignot and Westerhoff (2025)} \\ 
\scriptsize $\rho$ & \scriptsize $4.5$ & {\scriptsize So
domestic and foreign traders respond similarly} \\ 
\scriptsize $w^{flex}$ & \scriptsize $0.1$ & {\scriptsize Only 10\%
of firms revise their investment plans every period} \\ 
\scriptsize $\beta$ & \scriptsize $0.1$ & {\scriptsize Felipe and
Lanzafame (2020)} \\ 
\scriptsize $\Omega$ & \scriptsize 0.01 & {\scriptsize %
Westerhoff (2012)} \\ 
\scriptsize $\theta$ & \scriptsize $0.3$ & {\scriptsize Chaboud et al.
(2023)} \\ 
\scriptsize $\pi$ & \scriptsize $2$ & {\scriptsize Cimoli et al.
(2010), Bagnai et al. (2016), Felipe and Lanzafame (2020)} \\ \scriptsize $\Delta z^{NS}$ & \scriptsize $0.00003$ & {\scriptsize Compatible with a 1\% per capita annual output growth rate.} \\
\scriptsize $w^{F}$ & \scriptsize $0.75-1$ & {\scriptsize Strong fundamentalist majority in trading strategies.} \\\hline\hline
\end{tabular} 
\end{table}

To provide a more concrete view of the model's properties and gain additional economic intuition into what is happening, we calibrate it using the parameter values or ranges listed in Table \ref{parameters}. Their number is modest, no more than 10. Two play a critical role in generating endogenous persistent fluctuations and are related to the structure of the speculative sector. Where possible, we drew on estimates from the related literature. Still, given that our goal is not to match a specific real-world economy, our calibration strategy serves primarily as an approximation for illustrative purposes. Accordingly, our simulations should be viewed as qualitative in nature, aimed at capturing the key stylised facts outlined in the first part of the paper. For completeness, each figure specifies the parameter values employed in its generation.

For example, \href{#Mignot and Westerhoff 2025}{Mignot and Westerhoff (2025)} provide a set of numerical experiments matching the statistical properties of exchange rates in the European Union, the United Kingdom, and Japan. We use similar magnitudes for the parameters capturing the reaction of FX speculative traders, allowing $\mu \in (0,15)$ and fixing $\rho=4.5$. Regarding the reaction of capital accumulation to deviations of output growth with respect to the market clearing rate, we adopt conservative values for the share of firms that can adjust their investment plans, $w^{flex}=0.1$, and a lower speed coefficient $\beta=0.1$ in line with \href{#Felipe and Lanzafame 2020}{Felipe and Lanzafame (2020)} estimates. Evidence from \href{#Chaboud et al 2023}{Chaboud et al. (2023)} indicates that over the past 20 years, the share of global FX trading conducted with non-financial customers has declined from about 20\% to less than 10\%. Given that in developing countries such as Brazil or Peru, this share is higher, we set $\theta=0.3$. The values for the income elasticity of imports come from the dynamic trade-multiplier literature. Finally, we consider the fundamentalist heuristic as the absolute majority in trading strategies but allow some space for variation so that $w^F \in (0.75, 1)$.

\begin{figure}[tbp]
  \centering
  \subfloat[]{\includegraphics[width=3.1in]{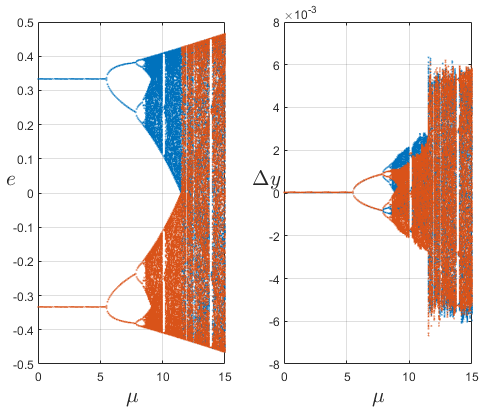}} \hspace{0.75cm}
  \subfloat[]{\includegraphics[width=3in]{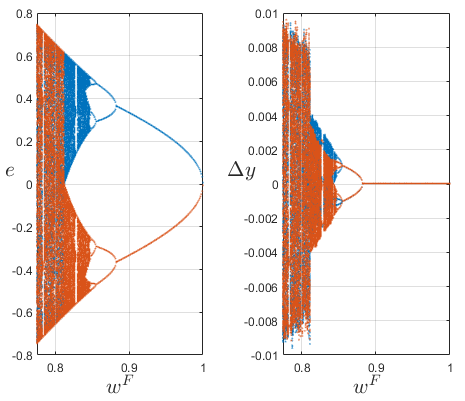}} \\
  \caption{Stronger reactions to FX deviations from the expected fundamental and weaker fundamentalists leading to instability. Orange (blue) colours indicate the equilibrium point $P_2$ ($P_3$) and the persistent dynamics around it. Parameters $\rho=4.5$, $w^{flex}=0.1$, $\beta=0.1$, $\Omega=0.01$, $\theta= 0.3$, $\pi=2$, $\Delta y^{BP}=0.00003$. Panel (a) uses $w^F=0.9$ and $w^C=0.1$. Panel (b) sets $\mu=4$.}\label{Flip}
\end{figure}

In Propositions 1-3, we presented the conditions under which one or multiple equilibria exist and the parameter set required for a Flip bifurcation to occur. Fig. \ref{Flip} confirms these findings. Focusing on panel (a), we fix the share of chartists at 10\%, so the model has three solutions. When the response of domestic speculators to exchange rate deviations from the fundamentals is relatively low, two of those equilibria are stable. We mark in blue the one with an undervalued FX, while in orange, the symmetrically overvalued case. As we increase $\mu$, a period-doubling bifurcation occurs; two chaotic attractors emerge and eventually merge. Panel (b) reports a similar bifurcation diagram fixing $\mu=4$ while increasing the share of fundamentalists from 0.75 to 1. They bring stability to the system, and in the limit, for $w^F=1$, there is a unique equilibrium value for the exchange rate.

\begin{figure}[tbp]
    \centering
    \includegraphics[width=0.6\linewidth]{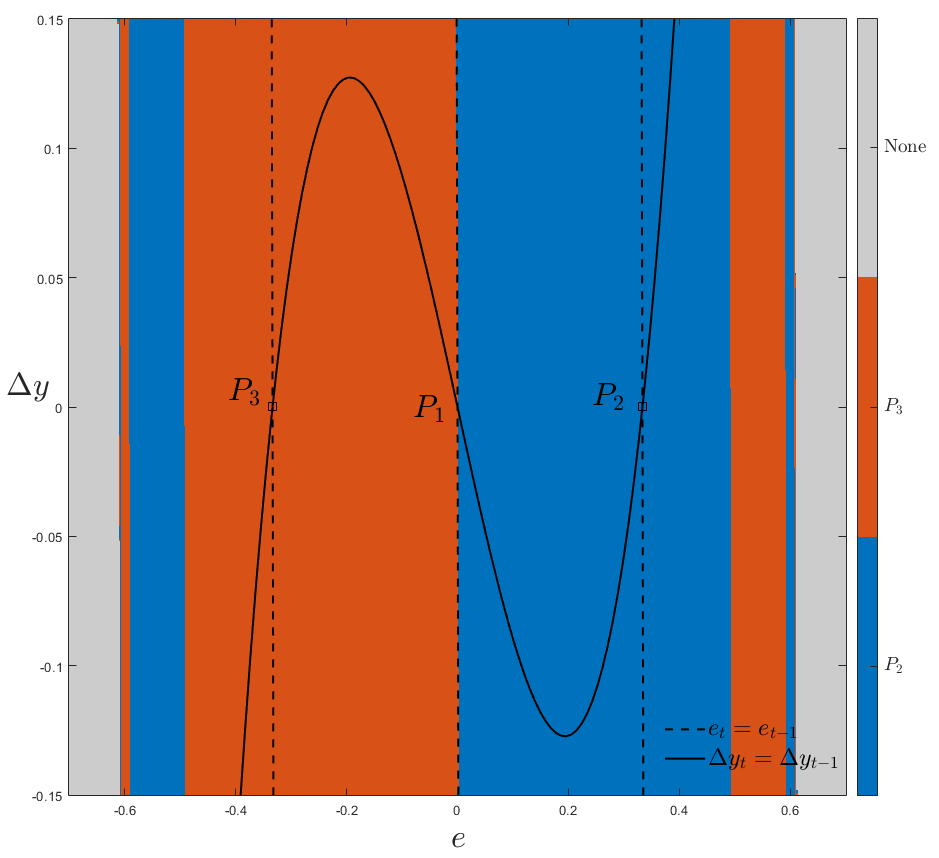}
    \caption{Discontinuous basins of attraction. Parameters $\mu=4.5$, $\rho=4.5$, $w^{flex}=0.1$, $\beta=0.1$, $\Omega=0.01$, $\theta= 0.3$, $\pi=2$, $\Delta y^{BP}=0.00003$, $w^F=0.9$, and $w^C=0.1$.}
    \label{Basins}
\end{figure}

The presence of multi-stability opens the door for the study of the corresponding basins of attraction.\footnote{A multistable dynamical system is one that possesses multiple stable equilibrium states or attractors. They might have periodic or aperiodic features. The system's long-term behaviour depends sensitively on initial conditions, with trajectories settling into different stable states depending on their starting point. In economics, this property provides a representation of path dependence and the idea that history matters.} Fig. \ref{Basins} indicates, in blue, initial conditions converging to $P_2$. In orange, we mark those leading to $P_3$. Stability properties are only local, and grey colours depict the region outside the two attracting boundaries. An interesting feature of the model is the presence of multiple disconnected regions. The discontinuity of the basins of attraction is a well-known attribute of non-invertible maps, and it may also have a non-trivial economic interpretation. Suppose an economy is in the neighbourhood of $P_2$. Small exogenous foreign exchange depreciation shocks may perturb the system, but only temporarily. However, if the depreciation is sufficiently large, it may lead to the orange region on the right, resulting in a jump to $P_3$. A similar transition works in the opposite direction. An economy in the neighbourhood of $P_3$ could unexpectedly jump to $P_2$ after a sufficiently strong currency appreciation.\footnote{In a non-invertible map like ours, the basins of attraction of coexisting attracting sets are generally not connected. It has such a structure because the total basin is obtained by taking all the preimages of any rank of the `immediate basin'. The boundary of divergent trajectories is given by the stable set of a saddle 2-cycle. The border between the two basins is given by the stable set of the fixed point in $P_1$ and all its preimages of any rank that have the saddle 2-cycle as a limit set. As we move along the horizontal axis to the right of $P_2$ or the left of $P_3$, the alternating orange and blue areas are infinitely many and accumulate monotonically toward the separatrix between the stable and unstable regions.}

Considering the relatively recent development experience of LA countries, we can only speculate whether a discontinuous basin of attraction is part of the explanation for what happened in the region during the 1990s. Several scholars have argued that LA has a more appreciated exchange rate with potentially more profound macro-development consequences (e.g. \href{#Rodrik 2008}{Rodrik, 2008}; \href{#Ugurlu and Razmi 2023}{Ugurlu and Razmi, 2023}; for an overview of the literature, see \href{#Demir and Razmi 2021}{Demir and Razmi, 2021}). While we do not argue this is the main explanation, the Asian crisis of 1997 indeed led to several substantial exchange rate depreciations, with countries such as Brazil abandoning fixed exchange regimes that had become unsustainable. In light of the model presented here, a sufficiently strong increase in $e$ might have paradoxically enabled these economies to reach the attracting region of the overvalued FX rate.

\begin{sidewaysfigure}[p]
\centering
  \includegraphics[width=8in]{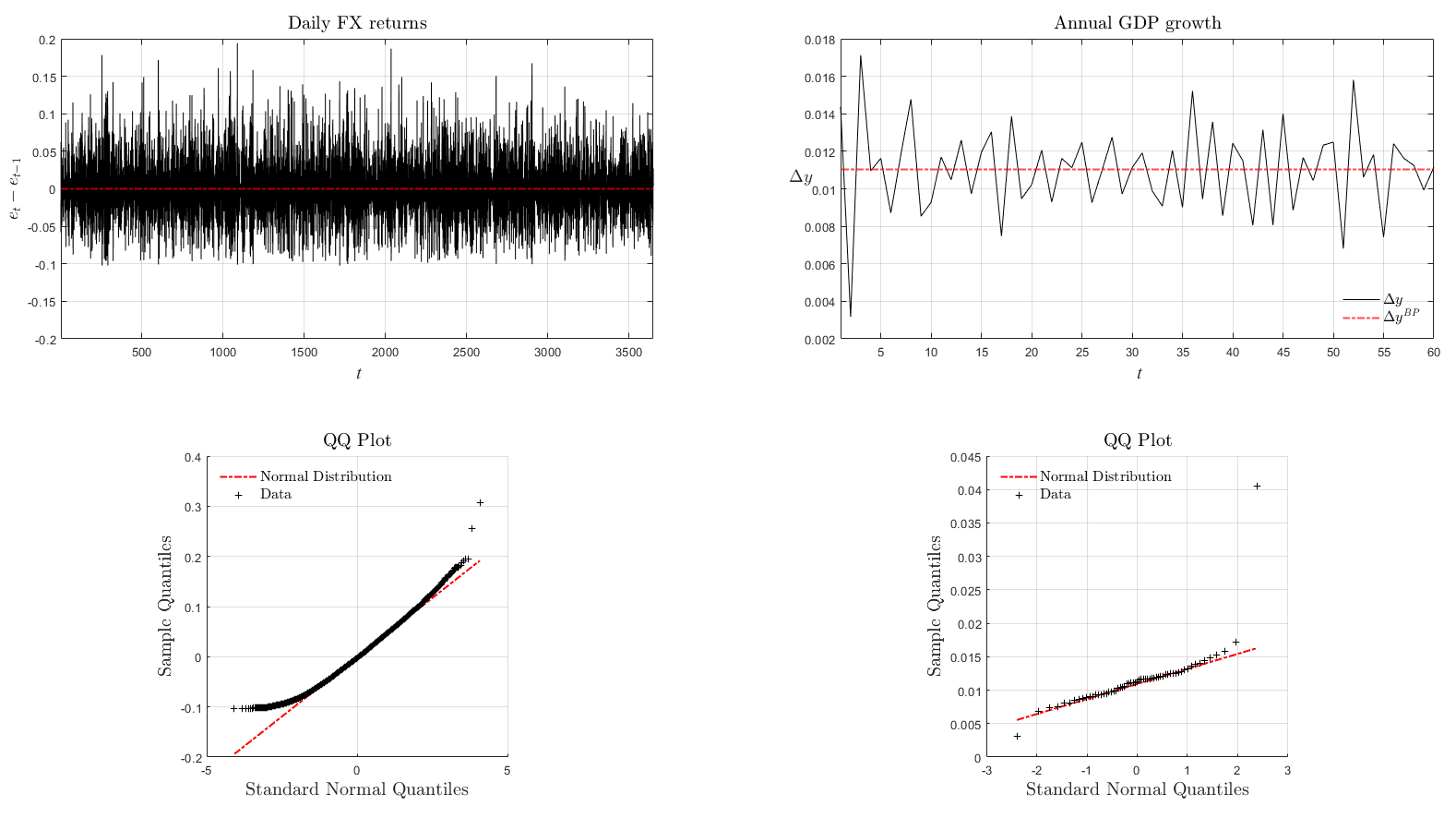} \bigskip
  \caption{Daily FX and annual output growth rates. Simulations based on the stochastic version of the model before the Flip bifurcation. Parameters $\mu=3.5$, $\rho=4.5$, $w^{flex}=0.1$, $\beta=0.1$, $\Omega=0.01$, $\theta= 0.3$, $\pi=2$, $\Delta y^{BP}=0.00003$, $w^F=0.9$, $w^C=0.1$. Anderson-Darling tests indicate that the series do not follow a normal distribution.}\label{Stochastic}
\end{sidewaysfigure}

Behavioural finance provides a valuable framework for studying foreign exchange, including `puzzles' such as their excess volatility and fat-tailed distributions. To assess whether our model is compatible with such properties, we proceed in two steps. First, we run the model for a combination of parameters for which we know $P_2$ and $P_3$ are stable, but allow for stochastic shocks. Fig. \ref{Stochastic} reports our main findings. Recall that our model is specified in such a way that $t$ is measured in terms of days. Therefore, to obtain series comparable with those in Fig. \ref{fig1}, we report daily FX returns and annual GDP growth rates. A visual inspection hints that they are very similar. The Anderson-Darling test indicates series do not follow a normal distribution. However, QQ plots suggest that the exchange rate deviations from the bell curve are not similar to those observed in the data and presented in the first part of the paper.

\begin{sidewaysfigure}[p]
\centering
  \includegraphics[width=8in]{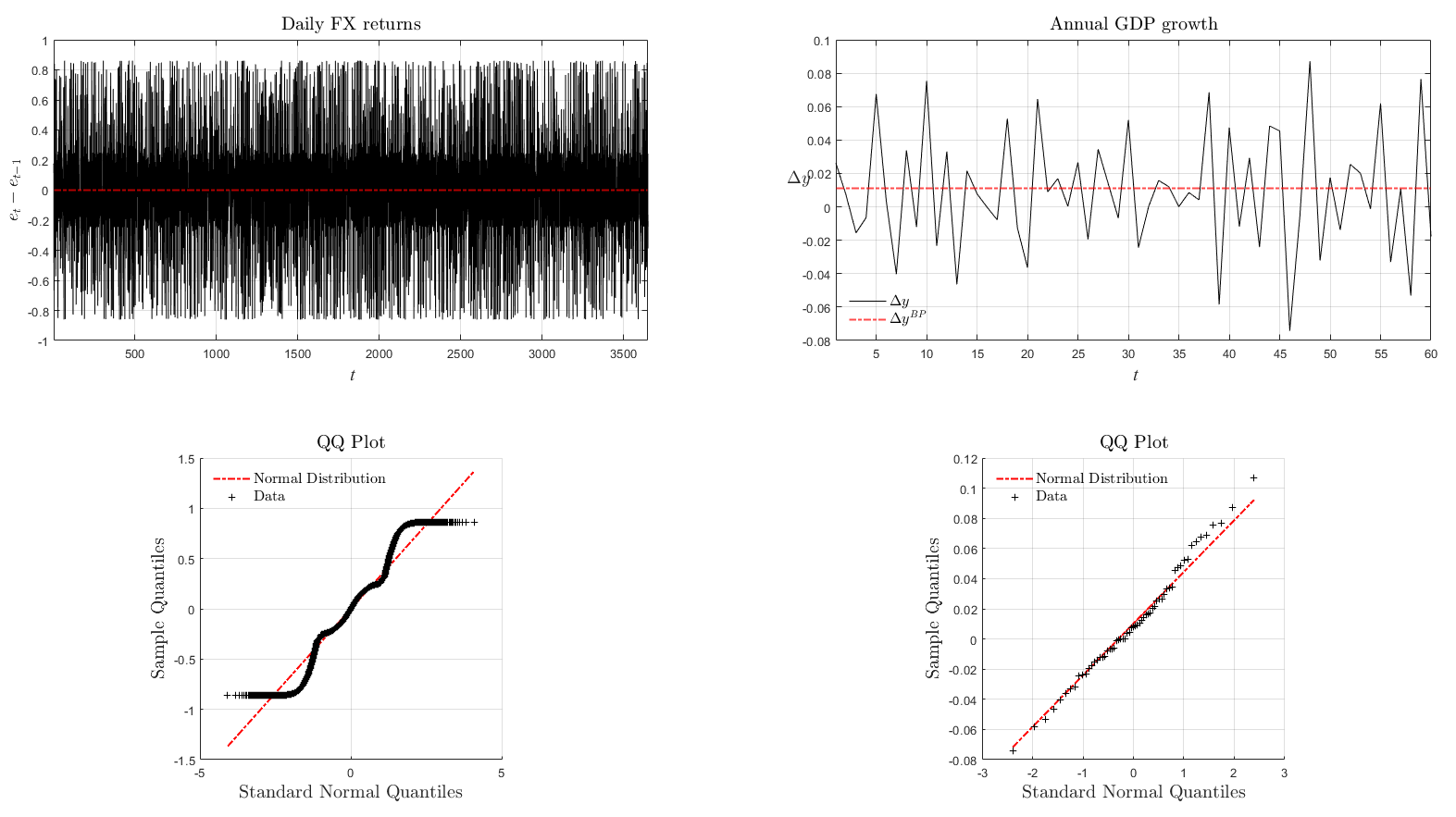} \bigskip
  \caption{Daily FX and annual output growth rates. Simulations based on the deterministic skeleton of the dynamic system after the Flip bifurcation. Parameters $\mu=14.99$, $\rho=4.5$, $w^{flex}=0.1$, $\beta=0.1$, $\Omega=0.01$, $\theta= 0.3$, $\pi=2$, $\Delta y^{BP}=0.00003$, $w^F=0.9$, $w^C=0.1$. Anderson-Darling test indicates series do not follow a normal distribution.}\label{Deterministic}
\end{sidewaysfigure}

If, instead, we take a parameter setting after the Flip bifurcation occurs, removing the stochastic component and relying only on the deterministic skeleton, a similar picture emerges. Fig. \ref{Deterministic} shows that fluctuations have a higher amplitude for both exchange and output growth rates. The model now also reproduces the possibility of economic recessions. Economic activity fluctuates around the trade-multiplier, and FX returns seem to have a zero mean. While the Anderson-Darling test suggests that none of them follow a normal distribution, as evident in the QQ plots, the deviations from the bell curve do not have the correct sign. 
The inclusion of trend-extrapolators will prove to be critical for this task.

\subsection{Introducing trend-extrapolators}

We are ready to introduce trend-extrapolation as a third trading strategy. Their demand is given by:
\begin{equation*}
    \Delta d^{E_i}_t=\mu \left( e_{t-1}- e_{t-2} \right)
\end{equation*}
so that they only look at the direction of the ongoing FX trajectory. For instance, if the exchange rate is depreciating, they project this process will continue in the foreseeable future, demanding more USDs. Inversely, they interpret that an ongoing appreciation process will persist in the next period.

If a similar heuristic is present among the suppliers of foreign currency, then it follows:
\begin{equation*}
    \Delta z^{E_i}_t= -\rho \left( e_{t-1}- e_{t-2} \right)
\end{equation*}
where the negative sign captures the fact that agents supplying foreign currency are non-nationals.

After some algebraic manipulations, it is easy to see that the new dynamic system is a 2D second-order nonlinear map:
\begin{align}
e_{t}  &  =e_{t-1}+(\mu+\rho)\left[  w^{F}\left(  -\Omega\Delta y_{t-1}%
-e_{t-1}\right)  ^{3}+w^{C}\left(  e_{t-1}+\Omega\Delta y_{t-1}\right)
+w^{E}\left(  e_{t-1}-e_{t-2}\right)  \right] \nonumber\\ \label{3D} \\
\triangle y_{t}  &  =\triangle y_{t-1}+w^{flex}\beta\left\{  \triangle
y^{BP}-\gamma\left[  w^{F}\left(  -\Omega\Delta y_{t-1}-e_{t-1}\right)
^{3}+w^{C}\left(  e_{t-1}+\Omega\Delta y_{t-1}\right)  \right.  \right.
\nonumber\\
& +\left. \left.  w^{E}\left(  e_{t-1}-e_{t-2}\right)\right]  -\triangle y_{t-1}\right\}
\nonumber
\end{align}
with similar equilibrium conditions as in (\ref{EqCond}).

Thus, we can state and prove the following Propositions regarding the existence of unique or multiple equilibria as well as the corresponding local stability properties.
\begin{proposition}
When there are only fundamentalists and extrapolators in the economy, i.e. $w^C = 0$, the dynamic system (\ref{3D}) admits $P_1$ as the unique equilibrium solution. The presence of chartists implies the existence of $P_2$ and $P_3$.
\label{prop 4}

\bigskip

\begin{proof}
See Appendix B.4.
\end{proof}

\end{proposition}


\begin{proposition}
When all three trading strategies can be found in the FX market, $P_1$ is unstable, whereas $P_{2,3}$ are locally stable in the region of the parameter space defined by:
\begin{align*}
    1-\Omega w^{E}\left(1+\gamma\right) & >0  \\
A+A_1 \left( w^{E} \right) & >0 \\
B+B_1 \left( w^{E} \right) & >0
\end{align*}
where $A_1 (\cdot)$ and $B_1(\cdot)$ are linear and quadratic functions in $w^E$, respectively. If a change in one of the parameters results in a violation of just the first condition, a Fold bifurcation occurs. If only the second inequality is not satisfied, a Flip bifurcation occurs. A violation of the third condition is associated with a Neimark-Sacker bifurcation.
\label{prop 5}

\bigskip

\begin{proof}
See Appendix B.5.
\end{proof}

\end{proposition}

\medskip

\begin{figure}[tbp]
  \centering
  \subfloat[]{\includegraphics[width=3in]{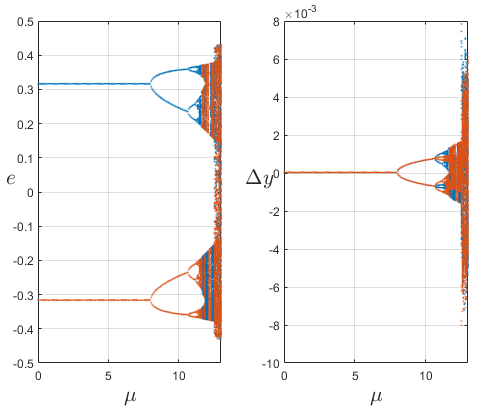}} \hspace{1cm}
  \subfloat[]{\includegraphics[width=2.9in]{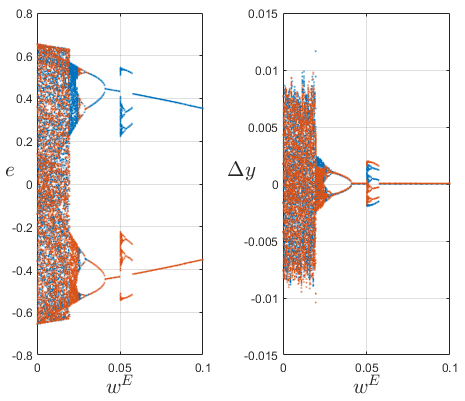}} \\
  \caption{Stronger reactions to FX deviations from the expected fundamental when there are extrapolators.  Instability is postponed, but the collapse point is anticipated. Orange (blue) colours indicate the equilibrium point $P_2$ ($P_3$) and the persistent dynamics around it. Parameters $\rho=4.5$, $w^{flex}=0.1$, $\beta=0.1$, $\Omega=0.01$, $\theta= 0.3$, $\pi=2$, $\Delta y^{BP}=0.00003$. Panel (a) uses $w^F=0.9$, $w^C=0.09$, and $w^E=0.01$. Panel (b) sets $w^F=0.8$ and $\mu=4$.}\label{Flip2}
\end{figure}

Accounting for trend-extrapolation does not change our main results in terms of the number of equilibria and the emergence of endogenous persistent fluctuations. Still, Fig. \ref{Flip2} suggests they create an interesting paradox. Looking at the panel (a) and comparing to Fig. \ref{Flip}, we need a much higher $\mu$ for the period-doubling cascade to begin. This result follows from the fact that trend-extrapolation smooths the contrast between fundamentalists' and chartists' strategies, increasing the stability region of the parameter space. However, they break the aperiodic attractor for much lower values of $\mu$, leading to explosive dynamics. Panel (b) fixes $w^F=0.8$ and gradually substitutes chartists with extrapolators until each group correspond to 10\% of the trading strategies. The paradox is clearer now as $w^E$ seems to stabilise the system. Yet, when $w^E>wC$, we just obtain explosive dynamics.

\begin{figure}[tbp]
    \centering
    \includegraphics[width=0.6\linewidth]{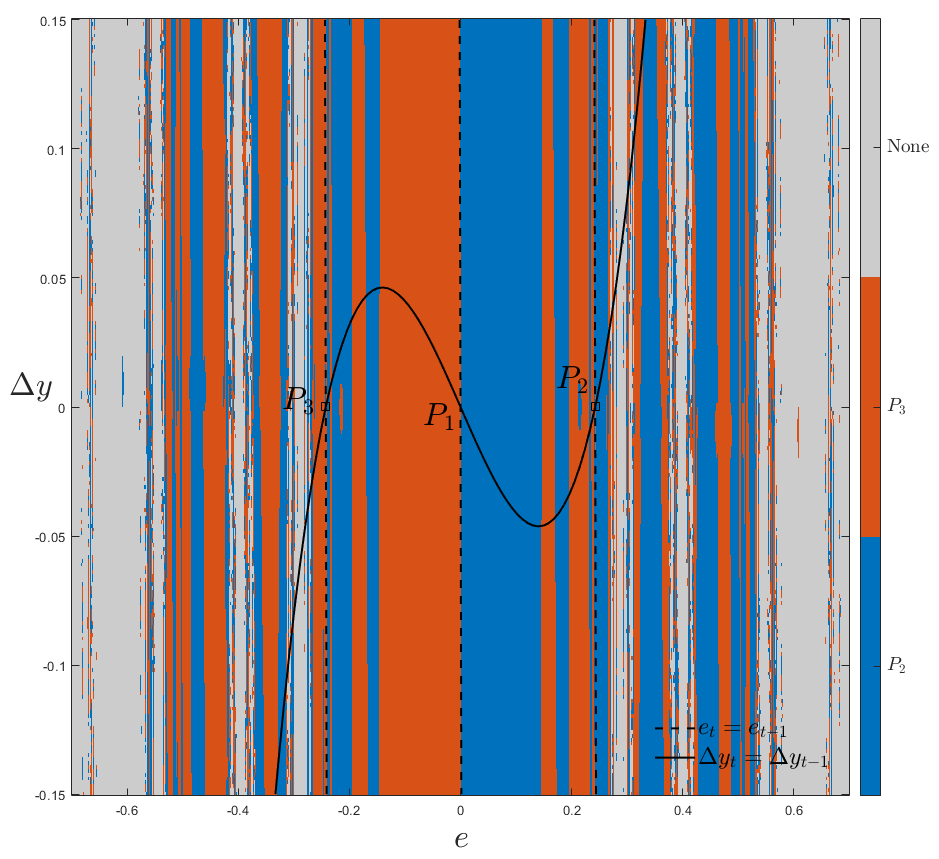}
    \caption{Discontinuous basins of attraction. Parameters $\mu=4$, $\rho=4.5$, $w^{flex}=0.1$, $\beta=0.1$, $\Omega=0.01$, $\theta= 0.3$, $\pi=2$, $\Delta y^{BP}=0.00003$, $w^F=0.85$, $w^C=0.05$, and $w^E=0.1$.}
    \label{Basins2}
\end{figure}

Fig. \ref{Basins2} reports the basins of attraction when we allow for 10\% of trend-extrapolators in the market. In principle, $P_{2,3}$ should be locally stable. Indeed, the blue colours indicate initial conditions converging to $P_2$, while those marked in orange are going to $P_3$. As before, the grey area corresponds to the initial conditions leading to explosive trajectories. The diagram shows another dimension of the instability brought by this third heuristic. We continue to have multiple disconnected regions. Yet, their number has greatly increased. For initial conditions slightly below $e_{PPP}$, there is convergence to the overvalued exchange rate solution. Analogously, for those just above $e_{PPP}$, the system converges to a more depreciated exchange rate. However, in the neighbourhood of those equilibria, minor shocks can result in dramatic shifts between attracting regions.\footnote{The external boundary between the stable and unstable regions in Fig. \ref{Basins2} continues to be shaped by a saddle-type two-cycle. With the introduction of trend-extrapolators, our current calibration implies that a contact bifurcation occurred, opening unstable holes in the basins of attraction. This is due to the merging of the local maximum and minimum with the saddle 2-cycle. After that contact, the maximum and the minimum have divergent trajectories, as well as all the points of a known interval around them and all the related preimages of any rank. This leads to alternating strips of basins with strips of divergent trajectories, which are very dangerous for points close to the local extrema and close to the saddle 2-cycle.}

The rationale for the aperiodic oscillations is the following. Suppose the FX is depreciated relative to the fundamental. As long as the deviation is small, chartists' demand for more foreign currency leads to a further devaluation of the exchange rate. Given that the speculative sector is increasingly absorbing the available foreign currency, there are fewer funds for non-speculative purposes. The real sector must adjust investment plans, thereby cooling output growth. The poor performance of the `real' economy is interpreted as a deterioration of economic fundamentals. Thus, agents expect a more depreciated FX fundamental, consolidating the initial deviation. Still, the gap with respect to the actual exchange rate remains open. Trend-extrapolators attenuate volatility but lead to more persistent deviations.

\begin{figure}[tbp]
    \centering
    \includegraphics[width=1\linewidth]{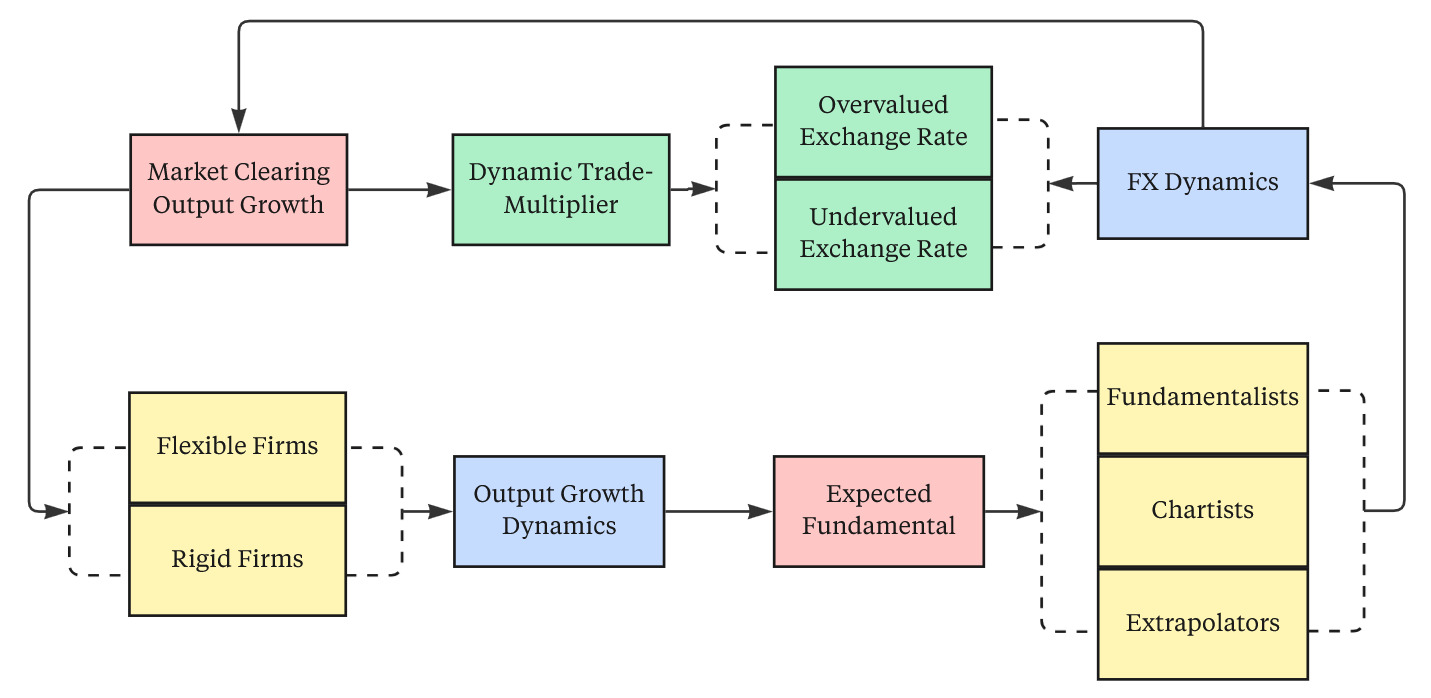}
    \caption{A summarising diagram. Blue indicates the endogenous variables in our map, $e$ and $\Delta y$. Red colours mark the connecting bridges between the financial and real spheres. Yellow corresponds to sources of heterogeneity both in financial and goods markets. Green shows the equilibria cases. Arrows indicate the direction of the relationship.}
    \label{Summary}
\end{figure}

When this gap becomes too large, fundamentalists' response dominates, and the demand for USDs by the speculative sector falls strongly. The result is an overvaluation of the domestic currency. At that point, more resources become available to the non-speculative sector. More flexible firms can increase their capital accumulation, thereby raising output growth. There is a perception that the economic fundamentals are improving, resulting in an appreciation of the expected fundamental rate. While it now consolidates the previous FX appreciation, this is not enough to close the gap. Chartists deepen the exchange rate overvaluation until it becomes too wide, triggering fundamentalists to buy in massive quantities. An exchange rate depreciation thus restarts the process. Fig. \ref{Summary} summarises the main transmission channels of the model.

\begin{sidewaysfigure}[p]
\centering
  \includegraphics[width=8in]{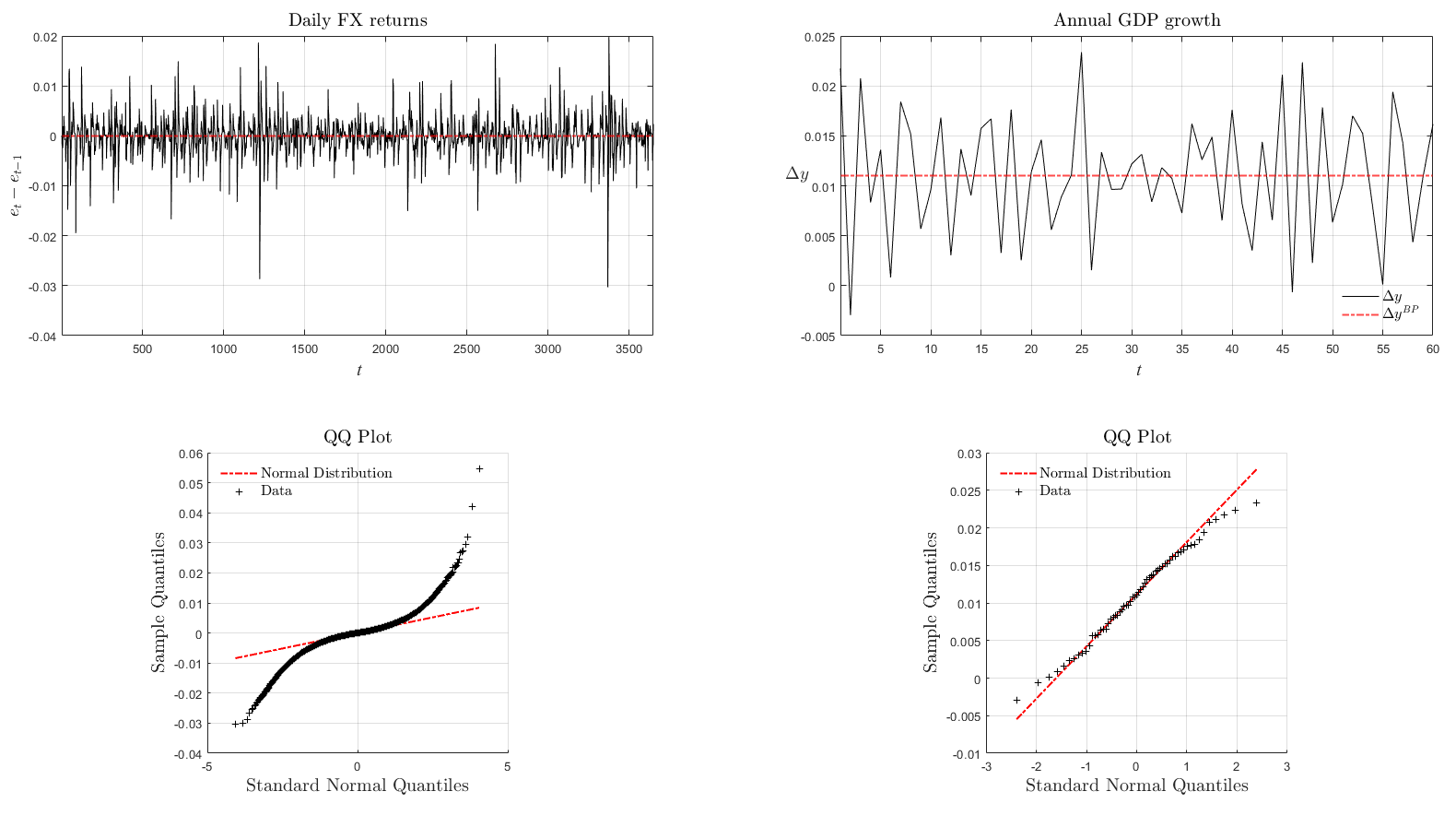} \bigskip
  \caption{Daily FX and annual output growth rates. Simulations based on the stochastic version of the model before the Flip bifurcation. Parameters $\mu=3.5$, $\rho=4.5$, $w^{flex}=0.1$, $\beta=0.1$, $\Omega=0.01$, $\theta= 0.3$, $\pi=2$, $\Delta y^{BP}=0.00003$, $w^F=0.9$, $w^C=0$, and $w^E=0.1$. Anderson-Darling test indicates series do not follow a normal distribution.}\label{Stochastic2}
\end{sidewaysfigure}

\begin{sidewaysfigure}[p]
\centering
  \includegraphics[width=8in]{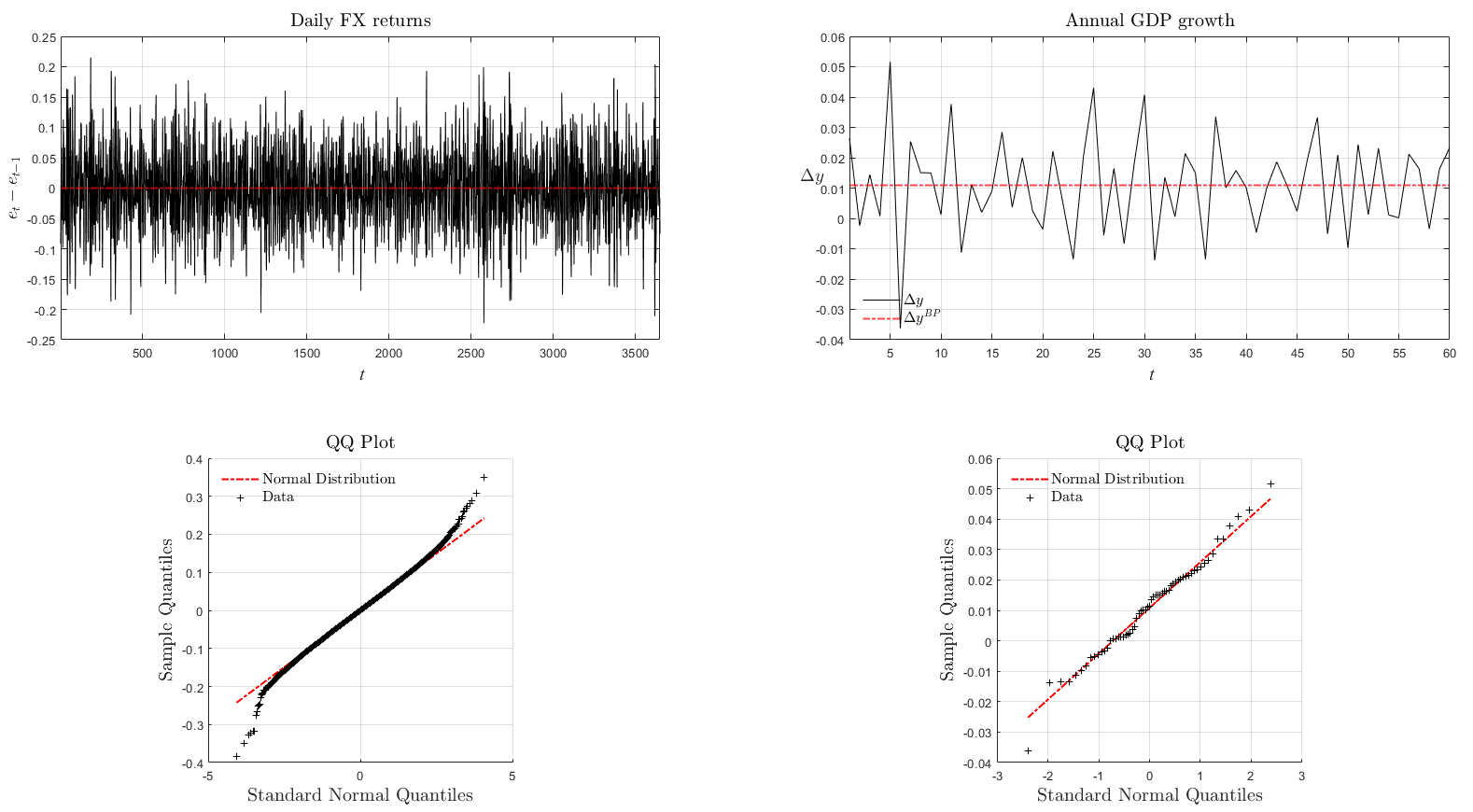} \bigskip
  \caption{Daily FX and annual output growth rates. Simulations based on the deterministic skeleton of the dynamic system after the Flip bifurcation. Parameters $\mu=8$, $\rho=4.5$, $w^{flex}=0.1$, $\beta=0.1$, $\Omega=0.01$, $\theta= 0.3$, $\pi=2$, $\Delta y^{BP}=0.00003$, $w^F=0.875$, $w^C=0.06$, and $w^E=0.065$. Anderson-Darling test indicates series do not follow a normal distribution.}\label{Deterministic2}
\end{sidewaysfigure}

As a final step, we evaluate whether the model continues to generate series that are not normally distributed. Recall that the model was designed with $t$ being measured in days. We use this series to calculate daily FX returns and the annual GDP growth rate. Our experiments demonstrate that extrapolators become a crucial component in enhancing the model's alignment with actual data. In Fig. \ref{Stochastic2}, we chose a set of parameters before the Flip bifurcation so that fluctuations are purely stochastic. In Fig. \ref{Deterministic2}, we mute the stochastic component but set parameters after the period-doubling cascade. In both cases, we obtain fat-tailed distributions comparable to those in Fig. \ref{fig1}. In fact, QQ plots reveal that exchange rate deviations with respect to the normal distribution have the right sign: inferior sample quantiles lie below those expected from the bell curve marked in red, while superior quantiles are above it.

\section{Final considerations}

The behavioural literature on the interplay between FX markets and the `real' economy has overlooked an important difference between developed and developing countries: the latter cannot trade in their national currencies. This poses a problem, as these economies require a constant flow of USDs to access capital and keep it running. To address this gap, we presented a novel heterogeneous agents model designed from the perspective of an emerging country that distinguishes between speculative and non-speculative sectors in the FX market. Its financial part took into account the existence of fundamentalism, chartism, and trend-extrapolation trading strategies.

We demonstrated that as long as non-speculative demand responds to domestic economic activity, a market-clearing output growth rate exists, which, in equilibrium, is equal to the ratio between the non-speculative FX supply growth and the income elasticity of demand for foreign assets. Numerical simulations suggest that the model is consistent with stylised facts on exchange rates and growth. These properties are more accurately represented in scenarios where all three types of speculators coexist. We provided a careful description of the economic intuition behind the obtained endogenous, persistent, and irregular fluctuations. Chartists appear as a source of instability in the model, while trend-extrapolators present a dual paradoxical effect. On the one hand, they stabilise the system by delaying the Flip bifurcation. On the other hand, if their share is too large, the system becomes more likely to explode, and the basin of attraction is significantly more discontinuous.

The discontinuity in the basins of attraction is a novel feature of our model, providing an innovative complementary explanation for the prevalence of overappreciated currencies in LA following the dramatic depreciations of the late 1990s. Small shocks are corrected without big jumps. However, larger depreciations (appreciations) can create a misalignment that triggers an overly strong run from fundamentalists, causing the economy to shift to a symmetrically overvalued (undervalued) foreign exchange equilibrium. In this context, extrapolators increase instability by creating strips of divergent trajectories.

Our parsimonious framework is flexible enough to be extended in several possible directions. The most evident is that our numerical experiments were mainly illustrative, and future research to improve them is to be encouraged. Another avenue is to properly address government actions, either directly through foreign exchange interventions or indirectly through monetary policy. US monetary policy adjustments have a significant impact on emerging economies, strongly influencing capital inflows and, consequently, exchange rates. A perennial question in the macro-development literature concerns whether a more depreciated currency leads to higher growth. The answer might depend on the behaviour of the FX market. Last but not least, proper treatment of different time scales between financial and real variables remains to be done, and contributions in that direction should be welcomed. These steps would allow us to get further insights into the possible connection channels with macro-development.

\FloatBarrier

\newpage

\appendix

\numberwithin{equation}{section}
\numberwithin{figure}{section}

\section{Empirical Appendix} 

\subsection{Exchange rates in the \textit{Andean Three}} \label{EmpiricAppendix}

\begin{figure}[h]
\centering
  \includegraphics[width=6.5in]{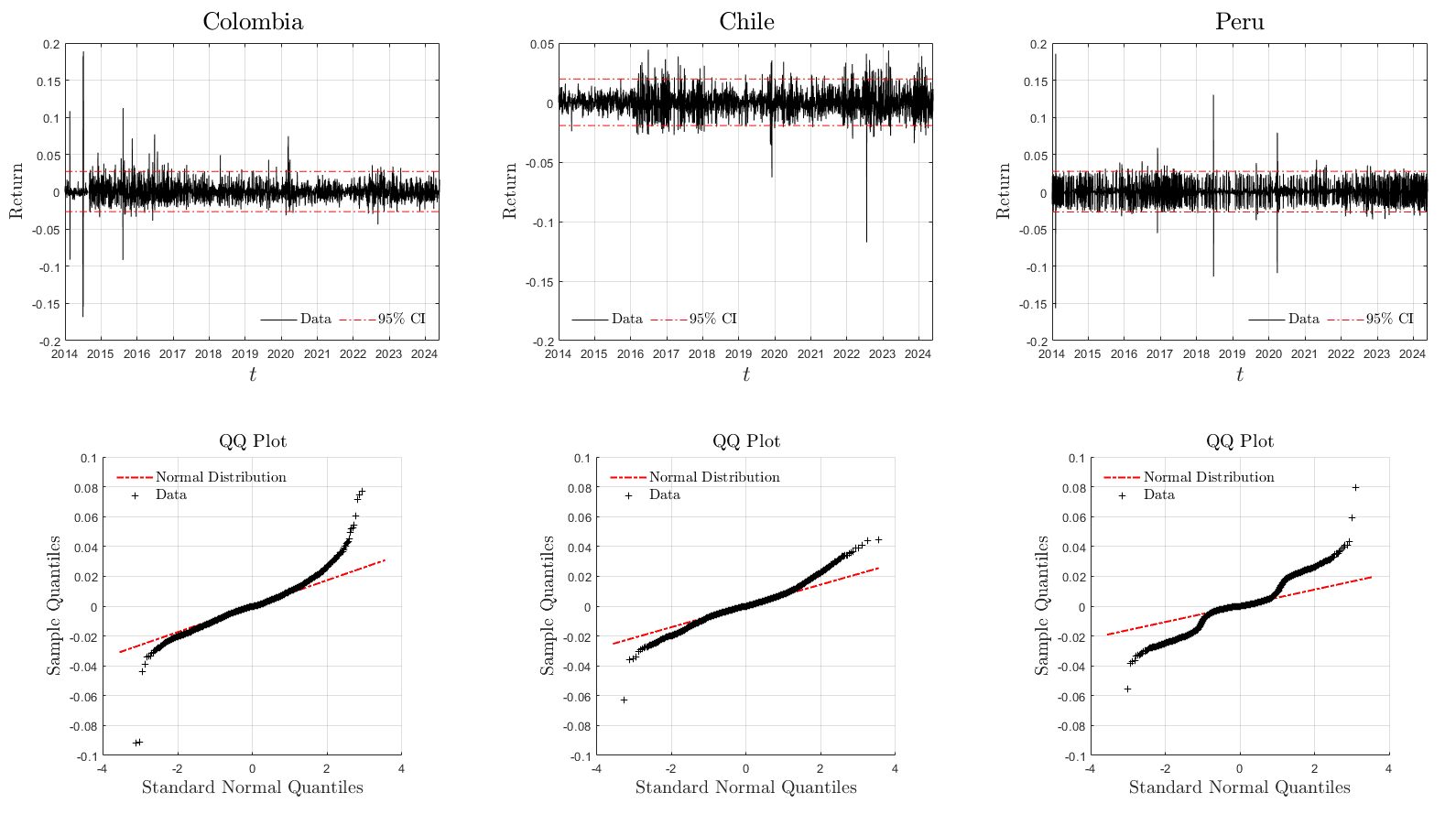} \bigskip
  \caption{High volatility and fat tails in exchange rates. Daily data for Colombia, Chile, and Peru, 2014-2024. Anderson-Darling test indicates series do not follow a normal distribution.}\label{fig2}
\end{figure}

Fig. \ref{fig2} reports the exchange rate returns for Colombia, Chile, and Peru. It shows that series have been extremely volatile over the past 10 years. The Anderson-Darling test confirms the visual reading of our QQ plots, indicating distributions are non-normal. Inferior sample quantiles lie below those expected from the bell curve; the opposite occurs in the superior quantiles. While is not our purpose to provide conclusive evidence on the matter, these simple plots provide initial insights aligned with the existing behavioural literature on FX rates in developed countries.

\subsection{Estimating the trade multiplier} \label{EmpiricAppendix2}

The dynamic trade-multiplier was conceived originally only for the non-speculative sector. Thus, the demand for USD comes from imports, while the supply of foreign currency is determined by exports. Suppose PPP holds. If imports grow at a rate proportional to GDP ($\pi \Delta y$) and exports grow at an exogenous rate ($\Delta z$), then the output growth rate that makes FX supply and demand grow at the same rate is given by:
\begin{equation}
    \Delta y_t^{BP}=\frac{\Delta z_t}{\pi_t} \label{Thirlwall}
\end{equation}
where $\pi>0$ is the income elasticity of demand for foreign assets or, in this context, just the income elasticity of imports.

This Appendix describes the data and estimation strategy used to obtain time-varying estimates supporting the empirical relevance of $\Delta y_t^{BP}$. For simplicity, we will rely on the aggregate version of the equation (for multisectoral considerations, see \href{#Araujo and Lima 2007}{Araújo and Lima, 2007}, and those afterwards). We rely on a Bayesian state space model (a literature review can be found in \href{#Chan and Strachan 2023}{Chan and Strachan, 2023}). We specify our system using two sets of equations: measurement and state.

Imports are given by the \textbf{\textit{measurement equation}}:
\begin{equation}
    m_{t}^{T} =\eta rer_{t}+\pi_{t}y_{t}^{T}+\varepsilon_{m,~t}  \label{StateSpaceM1}
\end{equation}
while $\pi_{t}$ is described by the \textbf{\textit{state equation}} that evolves over time:
\begin{equation}
 \pi_{t} =\pi_{t-1}+\varepsilon_{\pi,~t}   \label{StateSpaceM2}
\end{equation}
where $\eta$ is the price elasticity of imports, $rer$ is the real exchange rate, and $\varepsilon$ represents independent normally distributed errors with zero mean and constant variance. The superscript $T$ indicates we have removed short-run fluctuations using the Hodrick-Prescott (HP) filter with a smooth parameter $\lambda = 1600$. Given the small time dimension of our database and our lack of interest in price elasticities, $\eta$ does not change over time. We have the choice to report either filtered or smoothed estimates. As 
pointed out by \href{#Sims 2001}{Sims (2001)}, smoothed estimates provide insight into the difference between the best estimates made at time $t$ and ex-post estimates that utilise all available data today. We shall report them in what follows. 

\begin{figure}[tbp]
\centering
  \includegraphics[width=6.5in]{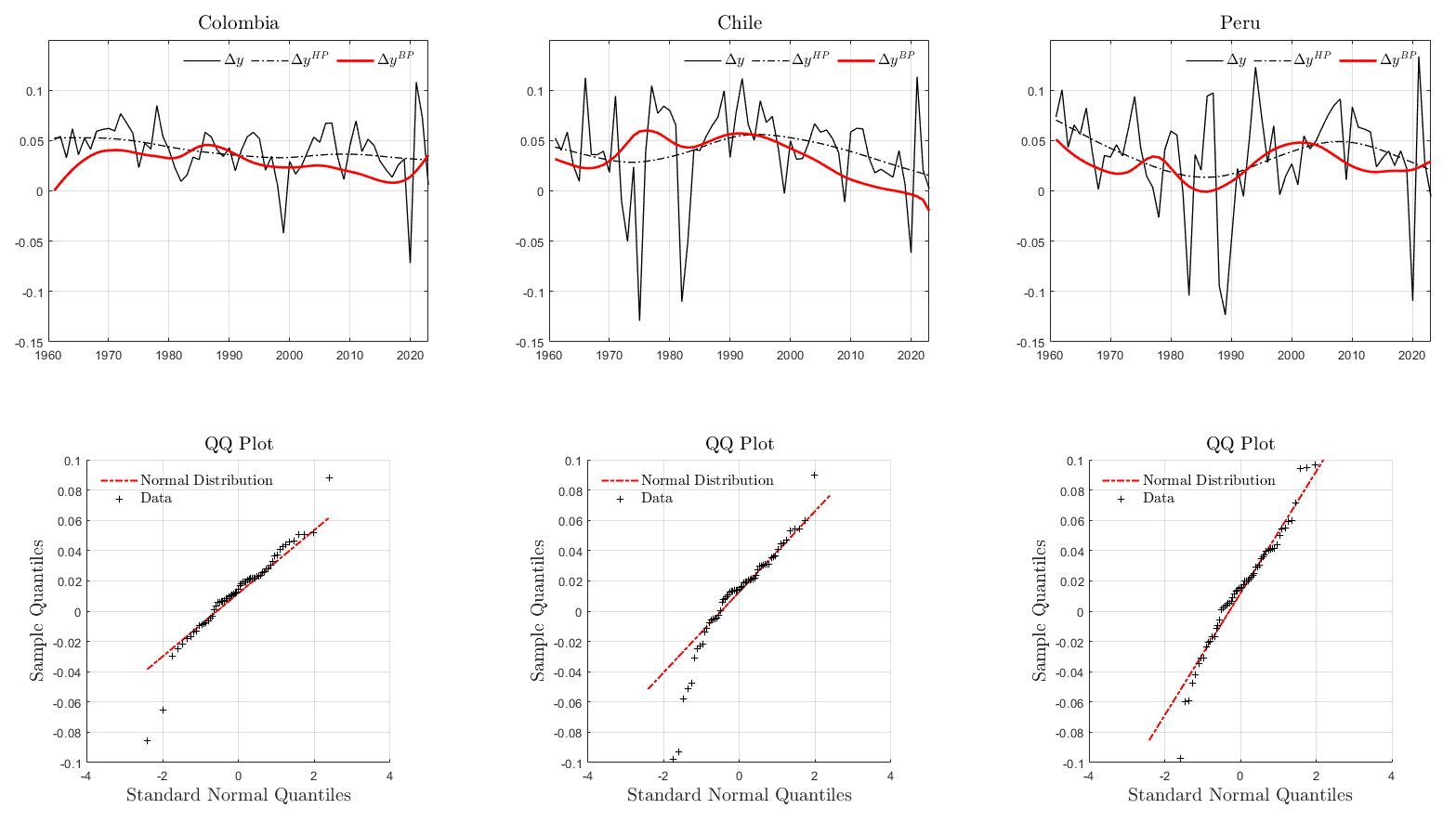} \bigskip
  \caption{Comparing actual growth rates ($\Delta y$) with the dynamic trade multiplier ($\Delta y^{BP}$). QQ plots indicate the difference $\Delta y - \Delta y^{BP}$ exhibits fat tails. Annual data for Colombia, Chile, and Peru, 1960-2023. Anderson-Darling test indicates series do not follow a normal distribution.}\label{fig4}
\end{figure}

Our data is annual for six major Latin American economies: Brazil, Mexico, Argentina, Colombia, Chile, and Peru. It comprehends the period 1960 to 2023. The output, exports, and import series are sourced from the World Development Indicators (WDI), measured in constant 2010 US dollars. Exports and imports include the value of merchandise, freight, insurance, transport, travel, royalties, license fees, and other services, such as communication, construction, financial, information, business, personal, and government services. They exclude compensation of employees, investment income, and transfer payments. Finally, the real exchange rate was obtained from the Bruegel dataset. All variables are defined in natural logarithms.

The rate of growth of exports was directly calculated from WDI and smoothed using the HP filter with $\lambda = 1600$. Once we estimate $\pi_{t}$, using Eqs. (\ref{StateSpaceM1}) and (\ref{StateSpaceM2}), we substitute both in Eq. (\ref{Thirlwall}), thus obtaining a time-varying estimate of the trade-multiplier. Our results are reported in Figs. \ref{fig3} and \ref{fig4}. In all six economies, $\Delta y_t^{BP}$ closely tracks actual growth rates that seem to fluctuate around it. They also move closer to simple HP detrended $\Delta y$. Furthermore, QQ plots suggest the difference $\Delta y_t - \Delta y_t^{BP}$ exhibits relatively high kurtosis and asymmetric skewness. This finding is confirmed by the Anderson-Darling test, which indicates series do not follow a normal distribution.

\FloatBarrier

\newpage

\section{Mathematical Appendix}

Recall our dynamic system:
\[
\left\{
\begin{array}
[c]{l}%
e_{t}=e_{t-1}+\left(  \mu+\rho\right)  \left[  w^{F}\left(  -e_{t-1}%
-\Omega\Delta y_{t-1}\right)  ^{3}+w^{C}\left(  e_{t-1}+\Omega\Delta
y_{t-1}\right)  \right] \\
\Delta y_{t}=\Delta y_{t-1}+w^{flex}\beta\left\{  \Delta y^{BP}-\gamma\left[
w^{F}\left(  -e_{t-1}-\Omega\Delta y_{t-1}\right)  ^{3}+w^{C}\left(
e_{t-1}+\Omega\Delta y_{t-1}\right)  \right]  -\Delta y_{t-1}\right\}
\end{array}
\right.
\]
with%
\[
0<\beta<1\text{, }\rho>0\text{, }\mu>0\text{, }0<\Omega<1\text{, }%
\gamma>0\text{, }0<w^{C}\text{, }w^{F}\text{, }w^{flex}<1
\]
The corresponding Jacobian matrix is given by:
\[
J=\left[
\begin{array}
[c]{cc}%
j_{11} & j_{12}\\
j_{21} & j_{22}%
\end{array}
\right]
\]%
where
\begin{align} \label{Jacobian}
j_{11}  & =1+\left(  \mu+\rho\right)  \left[  -3w^{F}\left(  -\bar{e}%
-\Omega\Delta\bar{y}\right)  ^{2}+w^{C}\right] \nonumber \\
j_{12}  & =\left(  \mu+\rho\right)  \left[  -3w^{F}\Omega\left(  -\bar
{e}-\Omega\Delta\bar{y}\right)  ^{2}+w^{C}\Omega\right]  \nonumber \\
j_{21}  & =w^{flex}\beta\left\{  -\gamma\left[  -3w^{F}\left(  -\bar{e}%
-\Omega\Delta\bar{y}\right)  ^{2}+w^{C}\right]  \right\} \\
j_{22}  & =1+w^{flex}\beta\left\{  -\gamma\left[  -3w^{F}\Omega\left(
-\bar{e}-\Omega\Delta\bar{y}\right)  ^{2}+w^{C}\Omega\right]  -1\right\} \nonumber
\end{align}
such that $\bar{e}=e_{t}=e_{t-1}$ and $\Delta\bar{y}=y_{t}=y_{t-1}$ are the steady state values of the two variables.

\subsection{Proof of Propositions 1}

We first study the two cases with no heterogeneity:

\begin{enumerate}
\item All agents are \textit{fundamentalists }$\left(  w^{F}=1\text{, }%
w^{C}=0\right)  $
\end{enumerate}

In this case, the dynamic system reduces to:
\[
\left\{
\begin{array}
[c]{l}%
e_{t}=e_{t-1}+\left(  \mu+\rho\right)  \left(  -e_{t-1}-\Omega\Delta
y_{t-1}\right)  ^{3}\\
\Delta y_{t}=\Delta y_{t-1}+w^{flex}\beta\left[  \Delta y^{BP}-\gamma\left(
-e_{t-1}-\Omega\Delta y_{t-1}\right)  ^{3}-\Delta y_{t-1}\right]
\end{array}
\right.
\]
with equilibrium conditions%
\[
\left\{
\begin{array}
[c]{l}%
0=\left(  \mu+\rho\right)  \left(  -\bar{e}-\Omega\Delta\bar{y}\right)  ^{3}\\
0=w^{flex}\beta\left[  \Delta y^{BP}-\gamma\left(  -\bar{e}-\Omega\Delta
\bar{y}\right)  ^{3}-\Delta\bar{y}\right]
\end{array}
\right.
\]
from which:%
\begin{align*}
\left(  -\bar{e}-\Omega\Delta\bar{y}\right)  ^{3}  & =0\\
\Delta y^{BP}-\Delta\bar{y}  & =0
\end{align*}
so that%

\[
P=\left(  \bar{e},\Delta\bar{y}\right)  =\left(  -\Omega\Delta y^{BP},\Delta
y^{BP}\right)
\]
and%

\begin{align*}
j_{11}  & =1-3\left(  \mu+\rho\right)  \left(  -\bar{e}-\Omega\Delta\bar
{y}\right)  ^{2}=1\\
j_{12}  & =3\left(  \mu+\rho\right)  \Omega\left(  -\bar{e}-\Omega\Delta
\bar{y}\right)  ^{2}=0\\
j_{21}  & =3w^{flex}\beta\gamma\left(  -\bar{e}-\Omega\Delta\bar{y}\right)
^{2}=0\\
j_{22}  & =1+w^{flex}\beta\left[  3\gamma\Omega\left(  -\bar{e}-\Omega
\Delta\bar{y}\right)  ^{2}-1\right]  =1-w^{flex}\beta
\end{align*}
\bigskip

Thus:%
\begin{align*}
\text{tr}J  & =2-w^{flex}\beta\\
\det J  & =1-w^{flex}\beta
\end{align*}
such that the local stability conditions become:%
\begin{align*}
\text{(i) } & 1+\text{tr}J+\det J   =3-2w^{flex}\beta>0\\
\text{(ii) } & 1-\text{tr}J+\det J   =1-2+w^{flex}\beta+1-w^{flex}\beta=0\\
\text{(iii) } & 1-\det J   =1-1+w^{flex}\beta=w^{flex}\beta>0
\end{align*}
Given that conditions (i) and (iii) are satisfied, having (ii) equal to zero
implies that either a Fold, Transcritical or Pitchfork bifurcation might occur.

\begin{enumerate}
\item[2.] All agents are \textit{chartists} $\left(  w^{F}=0\text{, }w^{C}=1\right)  $

Now, the dynamic system is given by:
\end{enumerate}%
\[
\left\{
\begin{array}
[c]{l}%
e_{t}=e_{t-1}+\left(  \mu+\rho\right)  \left(  e_{t-1}+\Omega\Delta
y_{t-1}\right) \\
\Delta y_{t}=\Delta y_{t-1}+w^{flex}\beta\left\{  \Delta y^{BP}-\gamma\left(
e_{t-1}+\Omega\Delta y_{t-1}\right)  -\Delta y_{t-1}\right\}
\end{array}
\right.
\]
with the unique equilibrium point as before
\[
P_{1}=\left(  \bar{e},\Delta\bar{y}\right)  =\left(  -\Omega\Delta
y^{BP},\Delta y^{BP}\right)
\]
The elements of the Jacobian matrix become:%
\begin{align*}
j_{11}  & =1+\mu+\rho\\
j_{12}  & =\left(  \mu+\rho\right)  \Omega\\
j_{21}  & =-\gamma w^{flex}\beta\\
j_{22}  & =1-w^{flex}\beta\left(  \gamma\Omega+1\right)
\end{align*}
such that%
\begin{align*}
\text{tr}J  & =1+\mu+\rho+1-w^{flex}\beta\left(  \gamma\Omega+1\right) \\
& =2+\mu+\rho-w^{flex}\beta\left(  \gamma\Omega+1\right) \\
\det J  & =\left(  1+\mu+\rho\right)  \left[  1-w^{flex}\beta\left(
\gamma\Omega+1\right)  \right]  +\left(  \mu+\rho\right)  \Omega w^{flex}%
\beta\gamma\\
& =1-w^{flex}\beta\left(  \gamma\Omega+1\right)  +\left(  \mu+\rho\right)
\left[  1-w^{flex}\beta\left(  \gamma\Omega+1\right)  \right]  +\left(
\mu+\rho\right)  \Omega w^{flex}\beta\gamma\\
& =1-w^{flex}\beta\left(  \gamma\Omega+1\right)  +\left(  \mu+\rho\right)
-\left(  \mu+\rho\right)  w^{flex}\beta\left(  \gamma\Omega+1\right)  +\left(
\mu+\rho\right)  \Omega w^{flex}\beta\gamma\\
& =1-w^{flex}\beta\left(  \gamma\Omega+1\right)  +\left(  \mu+\rho\right)
-\left(  \mu+\rho\right)  w^{flex}\beta\gamma\Omega-\left(  \mu+\rho\right)
w^{flex}\beta+\left(  \mu+\rho\right)  \Omega w^{flex}\beta\gamma\\
& =1-w^{flex}\beta\left(  \gamma\Omega+1\right)  +\left(  \mu+\rho\right)
\left(  1-w^{flex}\beta\right) \\
& =1-w^{flex}\beta\gamma\Omega-w^{flex}\beta+\left(  \mu+\rho\right)  \left(
1-w^{flex}\beta\right) \\
& =1-w^{flex}\beta+\left(  \mu+\rho\right)  \left(  1-w^{flex}\beta\right)
-w^{flex}\beta\gamma\Omega\\
& =\left(  1-w^{flex}\beta\right)  \left(  1+\mu+\rho\right)  -w^{flex}%
\beta\gamma\Omega
\end{align*}

The local stability conditions become:
\begin{align*}
\text{(i)} \quad &  1+\text{tr}J+\det J\\
& =1+2+\mu+\rho-w^{flex}\beta\left(  \gamma\Omega+1\right)  +\left(
1+\mu+\rho\right)  \left(  1-w^{flex}\beta\right)  -w^{flex}\beta\gamma
\Omega\\
& =1+2+\mu+\rho-w^{flex}\beta\left(  \gamma\Omega+1\right)  +1-w^{flex}%
\beta+\left(  \mu+\rho\right)  \left(  1-w^{flex}\beta\right)  -w^{flex}%
\beta\gamma\Omega\\
& =4+\mu+\rho-2w^{flex}\beta\left(  \gamma\Omega+1\right)  +\left(  \mu
+\rho\right)  \left(  1-w^{flex}\beta\right) \\
& =2\left(  2+\mu+\rho\right)  -2w^{flex}\beta\gamma\Omega-\left(  2+\mu
+\rho\right)  w^{flex}\beta\\
& =2\left(  2+\mu+\rho\right)  \left(  1-w^{flex}\beta\right)  -2w^{flex}%
\beta\gamma\Omega\gtreqqless0  \\ \\
\text{(ii)} \quad & 1-\text{tr}J+\det J\\
& =1-\left[  2+\mu+\rho-w^{flex}\beta\left(  \gamma\Omega+1\right)  \right] \\
& +\left(  1+\mu+\rho\right)  \left(  1-w^{flex}\beta\right)  -w^{flex}%
\beta\gamma\Omega\\
& =-w^{flex}\beta\left(  \gamma\Omega+1\right)  -\left(  1+\mu+\rho\right)
w^{flex}\beta-w^{flex}\beta\gamma\Omega\\
& =-\left(  \mu+\rho\right)  \rho w^{flex}\beta<0
\end{align*}

Given that (ii) is always violated, $P_{1}$ is unstable.

\subsection{Proof of Proposition 2}

We finally study the case with heterogeneity in FX $\left(
w^{F}\neq0\text{ and }w^{C}\neq0\right)$. Thus, we must consider the full dynamic system (\ref{2D}). The equilibrium points are found by solving:
\[
\left\{
\begin{array}
[c]{l}%
w^{F}\left(  -\bar{e}-\Omega\Delta\bar{y}\right)  ^{3}+w^{C}\left(  \bar
{e}+\Omega\Delta\bar{y}\right)  =0\\
\Delta y^{BP}-\gamma\left[  w^{F}\left(  -\bar{e}-\Omega\Delta\bar{y}\right)
^{3}+w^{C}\left(  \bar{e}+\Omega\Delta\bar{y}\right)  \right]  -\Delta\bar
{y}=0
\end{array}
\right.
\]
From the second equation:
\[
\Delta\bar{y}=\Delta y^{BP}%
\]
and inserting in the first:%
\[
\left(  -\bar{e}-\Omega\Delta y^{BP}\right)  \left[  w^{F}\left(  -\bar
{e}-\Omega\Delta y^{BP}\right)  ^{2}-w^{C}\right]  =0
\]
which is satisfied for%
\[
-\bar{e}-\Omega\Delta y^{BP}=0
\]
and%
\[
w^{F}\left(  -\bar{e}-\Omega\Delta\bar{y}\right)  ^{2}-w^{C}=0
\]
from which:%
\[
\bar{e}_{1}=-\Omega\Delta y^{BP}%
\]
and%
\[
\bar{e}_{2,3}=-\Omega\Delta y^{BP}\pm\sqrt{\frac{w^{C}}{w^{F}}}%
\]
Thus, the equilibrium points in this case are the following:%
\[
P_{1}=\left(  -\Omega\Delta y^{BP},\Delta y^{BP}\right)  \text{ \ \ }%
P_{2,3}=\left(  -\Omega\Delta y^{BP}\pm\sqrt{\frac{w^{C}}{w^{F}}},\Delta
y^{BP}\right)
\]

\subsection{Proof of Proposition 3}

To study local stability in this more general case, we must consider the elements of the Jacobian matrix as in \ref{Jacobian}. At $P_{1}=\left(  \bar{e}_{1},\Delta y^{BP}\right)  =\left(  -\Omega\Delta
y^{BP},\Delta y^{BP}\right)$, they become:
\begin{align*}
\left.  j_{11}\right\vert _{\left(  \bar{e}_{1},\Delta y^{BP}\right)  }  &
=1+\left(  \mu+\rho\right)  w^{C}\\
\left.  j_{12}\right\vert _{\left(  \bar{e}_{1},\Delta y^{BP}\right)  }  &
=\left(  \mu+\rho\right)  w^{C}\Omega\\
\left.  j_{21}\right\vert _{\left(  \bar{e}_{1},\Delta y^{BP}\right)  }  &
=-w^{flex}\beta\gamma w^{C}\\
\left.  j_{22}\right\vert _{\left(  \bar{e}_{1},\Delta y^{BP}\right)  }  &
=1-w^{flex}\beta\left(  \gamma w^{C}\Omega+1\right)
\end{align*}
such that:%
\begin{align*}
\text{tr}J  & =1+\left(  \mu+\rho\right)  w^{C}+1-w^{flex}\beta\left(  \gamma
w^{C}\Omega+1\right) \\
& =2+\left(  \mu+\rho\right)  w^{C}-w^{flex}\beta\left(  \gamma w^{C}%
\Omega+1\right)
\end{align*}
and%
\begin{gather*}
\det J=\left[  1+\left(  \mu+\rho\right)  w^{C}\right]  \left[  1-w^{flex}%
\beta\left(  \gamma w^{C}\Omega+1\right)  \right] \\
+\left(  \mu+\rho\right)  \Omega w^{C}\beta\gamma w^{flex}w^{C}\\
=1+\left(  \mu+\rho\right)  w^{C}-\left[  1+\left(  \mu+\rho\right)
w^{C}\right]  w^{flex}\beta\left(  \gamma w^{C}\Omega+1\right) \\
+\left(  \mu+\rho\right)  \Omega w^{C}\beta\gamma w^{flex}w^{C}\\
1+\left(  \mu+\rho\right)  w^{C}-w^{flex}\beta\left(  \gamma w^{C}%
\Omega+1\right)  -\left(  \mu+\rho\right)  w^{C}w^{flex}\beta\left(  \gamma
w^{C}\Omega+1\right) \\
+\left(  \mu+\rho\right)  \Omega w^{C}\beta\gamma w^{flex}w^{C}\\
=1+\left(  \mu+\rho\right)  w^{C}-w^{flex}\beta\left(  \gamma w^{C}%
\Omega+1\right)  -\left(  \mu+\rho\right)  w^{C}w^{flex}\beta
\end{gather*}

It is easy to verify that the second local stability condition is never
satisfied:%
\begin{align*}
& 1-\text{tr}J+\det J=1-2-\left(  \mu+\rho\right)  w^{C}+w^{flex}\beta\left(
\gamma w^{C}\Omega+1\right) \\
& +1+\left(  \mu+\rho\right)  w^{C}-w^{flex}\beta\left(  \gamma w^{C}%
\Omega+1\right)  -\left(  \mu+\rho\right)  w^{C}w^{flex}\beta\\
& =-\left(  \mu+\rho\right)  w^{C}w^{flex}\beta<0
\end{align*}
implying that $P_{1}$ is unstable.

Given that:
\[
\left(  -\bar{e}_{2,3}-\Omega\Delta y^{BP}\right)  ^{2}=\left(  \Omega\Delta
y^{BP}\mp\sqrt{\frac{w^{C}}{w^{F}}}-\Omega\Delta y^{BP}\right)  ^{2}%
=\frac{w^{C}}{w^{F}}%
\]
at $P_{2,3}=\left(  \bar{e}_{2,3},\Delta y^{BP}\right)  =\left(  -\Omega\Delta
y^{BP}\pm\sqrt{\frac{w^{C}}{w^{F}}},\Delta y^{BP}\right)  $, the elements of the Jacobian become:
\begin{align*}
\left.  j_{11}\right\vert_{\left(  \bar{e}_{2,3},\Delta y^{BP}\right)}  & =1-2\left(  \mu+\rho\right)  w^{C}\\
\left.  j_{12}\right\vert_{\left(  \bar{e}_{2,3},\Delta y^{BP}\right)}  & =-2\left(  \mu+\rho\right)  \Omega w^{C}\\
\left.  j_{21}\right\vert_{\left(  \bar{e}_{2,3},\Delta y^{BP}\right)}  & =2w^{flex}\beta\gamma w^{C}\\
\left.  j_{22}\right\vert_{\left(  \bar{e}_{2,3},\Delta y^{BP}\right)}  & =1+w^{flex}\beta\left\{  -\gamma\left[  -3w^{F}\Omega\left(
-\bar{e}_{2,3}-\Omega\Delta\bar{y}^{BP}\right)  ^{2}+w^{C}\Omega\right]
-1\right\} \\
& =1-w^{flex}\beta\left(  1-2\gamma\Omega w^{C}\right)
\end{align*}
such that%
\begin{align*}
\text{tr}J  & =1-2\left(  \mu+\rho\right)  w^{C}+1-w^{flex}\beta\left(
1-2\gamma\Omega w^{C}\right) \\
& =2\left[  1-\left(  \mu+\rho\right)  w^{C}\right]  -w^{flex}\beta\left(
1-2\gamma\Omega w^{C}\right) \\
\det J  & =\left[  1-2\left(  \mu+\rho\right)  w^{C}\right]  \left[
1-w^{flex}\beta\left(  1-2\gamma\Omega w^{C}\right)  \right] \\
& +4\left(  \mu+\rho\right)  \Omega w^{C}w^{flex}\beta\gamma w^{C}%
\end{align*}
from which:%
\begin{align*}
\text{(i)} \quad  & 1+\text{tr}J+\det J\\
& =1+2\left[  1-\left(  \mu+\rho\right)  w^{C}\right]  -w^{flex}\beta\left(
1-2\gamma\Omega w^{C}\right) \\
& +\left[  1-2\left(  \mu+\rho\right)  w^{C}\right]  \left[  1-w^{flex}%
\beta\left(  1-2\gamma\Omega w^{C}\right)  \right]  +4\left(  \mu+\rho\right)
\Omega w^{C}w^{flex}\beta\gamma w^{C}\\
& =1+2-2\left(  \mu+\rho\right)  w^{C}-w^{flex}\beta\left(  1-2\gamma\Omega
w^{C}\right) \\
& +1-2\left(  \mu+\rho\right)  w^{C}-\left[  1-2\left(  \mu+\rho\right)
w^{C}\right]  w^{flex}\beta\left(  1-2\gamma\Omega w^{C}\right)  +4\left(
\mu+\rho\right)  \Omega w^{C}w^{flex}\beta\gamma w^{C}\\
& =4-4\left(  \mu+\rho\right)  w^{C}-w^{flex}\beta\left(  1-2\gamma\Omega
w^{C}\right) \\
& -\left[  1-2\left(  \mu+\rho\right)  w^{C}\right]  w^{flex}\beta\left(
1-2\gamma\Omega w^{C}\right)  +4\left(  \mu+\rho\right)  \Omega w^{C}%
w^{flex}\beta\gamma w^{C}\\
& =4-4\left(  \mu+\rho\right)  w^{C}-w^{flex}\beta\left(  1-2\gamma\Omega
w^{C}\right) \\
& -w^{flex}\beta\left(  1-2\gamma\Omega w^{C}\right)  +2\left(  \mu
+\rho\right)  w^{C}w^{flex}\beta\left(  1-2\gamma\Omega w^{C}\right)
+4\left(  \mu+\rho\right)  \Omega w^{C}w^{flex}\beta\gamma w^{C}%
\end{align*}
\begin{align*}
& =4-4\left(  \mu+\rho\right)  w^{C}-w^{flex}\beta\left(  1-2\gamma\Omega
w^{C}\right) \\
& -w^{flex}\beta\left(  1-2\gamma\Omega w^{C}\right)  +2\left(  \mu
+\rho\right)  w^{C}w^{flex}\beta-4\left(  \mu+\rho\right)  w^{C}w^{flex}%
\beta\gamma\Omega w^{C}+4\left(  \mu+\rho\right)  \Omega w^{C}w^{flex}%
\beta\gamma w^{C}%
\end{align*}%
\[
=4-4\left(  \mu+\rho\right)  w^{C}-2w^{flex}\beta\left(  1-2\gamma\Omega
w^{C}\right)  +2\left(  \mu+\rho\right)  w^{C}w^{flex}\beta\gtreqqless0
\]%
\[
4-4\left(  \mu+\rho\right)  w^{C}+2 \left[ \left( \mu + \rho+ 2\gamma \Omega \right)w^C-1 \right]w^{flex}\gtreqqless0
\]%
\begin{align*}
& \text{(ii)} \quad 1-\text{tr}J+\det J\\
& =1-2\left[  1-\left(  \mu+\rho\right)  w^{C}\right]  +w^{flex}\beta\left(
1-2\gamma\Omega w^{C}\right) \\
& +\left[  1-2\left(  \mu+\rho\right)  w^{C}\right]  \left[  1-w^{flex}%
\beta\left(  1-2\gamma\Omega w^{C}\right)  \right]  +4\left(  \mu+\rho\right)
\Omega w^{C}w^{flex}\beta\gamma w^{C}\\
& =1-2\left[  1-\left(  \mu+\rho\right)  w^{C}\right]  +w^{flex}%
\beta-2w^{flex}\beta\gamma\Omega w^{C}\\
& +\left[  1-2\left(  \mu+\rho\right)  w^{C}\right]  \left[  1-w^{flex}%
\beta+2w^{flex}\beta\gamma\Omega w^{C}\right]  +4\left(  \mu+\rho\right)
\Omega w^{C}w^{flex}\beta\gamma w^{C}\\
& =2\left(  \mu+\rho\right)  w^{C}-2\left(  \mu+\rho\right)  w^{C}\left[
1-w^{flex}\beta+2w^{flex}\beta\gamma\Omega w^{C}\right]  +4\left(  \mu
+\rho\right)  \Omega w^{C}w^{flex}\beta\gamma w^{C}\\
& =2\left(  \mu+\rho\right)  w^{C}w^{flex}\beta>0
\end{align*}

\begin{align*}
& \text{(iii)} \quad 1-\det J\\
& =1-\left[  1-2\left(  \mu+\rho\right)  w^{C}\right]  \left[  1-w^{flex}%
\beta\left(  1-2\gamma\Omega w^{C}\right)  \right]  -4\left(  \mu+\rho\right)
\Omega w^{C}w^{flex}\beta\gamma w^{C}\\
& =1-1+w^{flex}\beta\left(  1-2\gamma\Omega w^{C}\right)  +2\left(  \mu
+\rho\right)  w^{C}\left[  1-w^{flex}\beta\left(  1-2\gamma\Omega
w^{C}\right)  \right] \\
& -4\left(  \mu+\rho\right)  \Omega w^{C}w^{flex}\beta\gamma w^{C}\\
& =w^{flex}\beta-2\gamma\Omega w^{C}w^{flex}\beta+2\left(  \mu+\rho\right)
w^{C}\left(  1-w^{flex}\beta+w^{flex}\beta2\gamma\Omega w^{C}\right) \\
& -4\left(  \mu+\rho\right)  \Omega w^{C}w^{flex}\beta\gamma w^{C}\\
& =w^{flex}\beta-2\gamma\Omega w^{C}w^{flex}\beta+2\left(  \mu+\rho\right)
w^{C}-2\left(  \mu+\rho\right)  w^{C}w^{flex}\beta\\
& =\left(  1-2\gamma\Omega w^{C}\right)  w^{flex}\beta+2\left(  \mu
+\rho\right)  w^{C}\left(  1-w^{flex}\beta\right)  \gtreqqless0
\end{align*}

Therefore, the equilibrium points $P_{2.3}$ are locally stable provided that:
\begin{align*}
    4-4\left(  \mu+\rho\right)  w^{C}-2w^{flex}\beta\left(  1-2\gamma\Omega
w^{C}\right)  +2\left(  \mu+\rho\right)  w^{C}w^{flex}\beta>0
\end{align*}
and
\begin{align*}
\left(  1-2\gamma\Omega w^{C}\right)  w^{flex}\beta+2\left(  \mu
+\rho\right)  w^{C}\left(  1-w^{flex}\beta\right)>0
\end{align*}
A violation of the first while the second is satisfied is associated with a Flip bifurcation. Moreover, if the second condition is violated while the first is satisfied, a Neimark-Sacker bifurcation occurs. Fig. \ref{Bifurcations} illustrates these two cases for our baseline calibration.

\begin{figure}[tbp]
    \centering
    \includegraphics[width=0.75\linewidth]{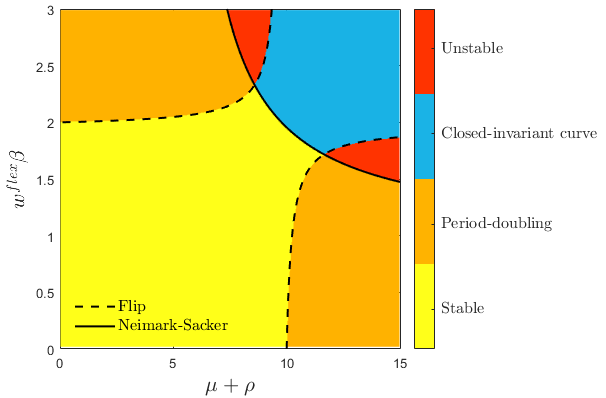}
    \caption{Flip and Neimark-Sacker Bifurcations. Parameters $\Omega=0.01$, $\theta=0.3$, $\pi=2$, $\Delta y^{BP}$, $w^F=0.9$, $w^C=0.1$. Notice that because $0<w^{flex} \beta <1$, only the Flip bifurcation is feasible, even though a Neimark-Sacker is in principle possible.}
    \label{Bifurcations}
\end{figure}

\subsection{Proof of Proposition 4}

When there are also trend extrapolators, our model becomes a 2-dimensional second-order nonlinear map (\ref{3D}). In steady state, $e_{t}=e_{t-1}=e_{t-2}$ and $\Delta y_{t}=\Delta y_{t-1}$, and the equilibrium conditions are:
\[
\left\{
\begin{array}
[c]{l}%
0=w^{F}\left(  -\bar{e}-\Omega\Delta\bar{y}\right)  ^{3}+w^{C}\left(  \bar
{e}+\Omega\Delta\bar{y}\right) \\
0=\Delta y^{BP}-\gamma\left[  w^{F}\left(  -\bar{e}-\Omega\Delta\bar
{y}\right)  ^{3}+w^{C}\left(  \bar{e}+\Omega\Delta\bar{y}\right)  \right]
-\Delta\bar{y}%
\end{array}
\right.
\]
from which we obtain the same points $P_{1}\ $and $P_{2,3}$ as before.

When in the economy there are only fundamentalists and extrapolators, so that
$w^{C}=0$, it is easy to see:
\[
\left\{
\begin{array}
[c]{l}%
\left(  -\bar{e}-\Omega\Delta\bar{y}\right)  ^{3}=0\\
\Delta y^{BP}-\Delta\bar{y}=0
\end{array}
\right.
\]
Thus, the system admits only the equilibrium point $P_{1}=\left(  -\Omega\Delta y^{BP},\Delta y^{BP}\right) $, whereas when there are also chartists it admits in addition the equilibrium points $P_{2,3}=\left(  -\Omega\Delta y^{BP}\pm\sqrt{\frac{w^{C}}{w^{F}}},\Delta y^{BP}\right)$.

\subsection{Proof of Proposition 5}

We rewrite the dynamic system (\ref{3D}) using the
auxiliary variable $z_{t}=e_{t-1}$ such that it becomes:%
\[
\left\{
\begin{array}
[c]{l}%
e_{t}=e_{t-1}+\left(  \mu+\rho\right)  \left[  w^{F}\left(  -e_{t-1}%
-\Omega\Delta y_{t-1}\right)  ^{3}+w^{C}\left(  e_{t-1}+\Omega\Delta
y_{t-1}\right)  +w^{E}\left(  e_{t-1}-z_{t-1}\right)  \right] \\
\Delta y_{t}=\Delta y_{t-1}+w^{flex}\beta\left\{  \Delta y^{BP}-\gamma\left[
w^{F}\left(  -e_{t-1}-\Omega\Delta y_{t-1}\right)  ^{3}+w^{C}\left(
e_{t-1}+\Omega\Delta y_{t-1}\right)  \right.  \right. \\
\left.  +w^{E}\left(  e_{t-1}-z_{t-1}\right)  ]-\Delta y_{t-1}\right\} \\
z_{t}=e_{t-1}%
\end{array}
\right.
\]

Thus, the Jacobian matrix of the dynamic system is given by:%
\[
J=\left[
\begin{array}
[c]{ccc}%
j_{11} & j_{12} & j_{13}\\
j_{21} & j_{22} & j_{23}\\
1 & 0 & 0
\end{array}
\right]
\]
where%
\begin{align*}
j_{11}  &  =1+\left(  \mu+\rho\right)  \left[  -3w^{F}\left(  -\bar{e}%
-\Omega\Delta\bar{y}\right)  ^{2}+w^{C}+w^{E}\right] \\
j_{12}  &  =\left(  \mu+\rho\right)  \left[  -3w^{F}\Omega\left(  -\bar
{e}-\Omega\Delta\bar{y}\right)  ^{2}+w^{C}\Omega\right] \\
j_{13}  &  =-\left(  \mu+\rho\right)  w^{E}\\
j_{21}  &  =-w^{flex}\beta\gamma\left[  -3w^{F}\left(  -\bar{e}-\Omega
\Delta\bar{y}\right)  ^{2}+w^{C}+w^{E}\right] \\
j_{22}  &  =1+w^{flex}\beta\left\{  -\gamma\left[  -3w^{F}\Omega\left(
-\bar{e}-\Omega\Delta\bar{y}\right)  ^{2}+w^{C}\Omega\right]  -1\right\} \\
&  =1-w^{flex}\beta\gamma\left[  -3w^{F}\Omega\left(  -\bar{e}-\Omega
\Delta\bar{y}\right)  ^{2}+w^{C}\Omega\right]  -w^{flex}\beta\\
j_{23}  &  =-w^{flex}\beta w^{E}\\
j_{31}  &  =1\\
j_{32}  &  =j_{33}=0
\end{align*}

This means that the characteristic equation can be written as:%
\[
\lambda^{3}+a\lambda^{2}+b\lambda+c=0
\]
where%
\[
a=-\text{tr}J=-j_{11}-j_{22}%
\]%
\[
b=\left\vert
\begin{array}
[c]{cc}%
j_{22} & j_{23}\\
0 & 0
\end{array}
\right\vert +\left\vert
\begin{array}
[c]{cc}%
j_{11} & j_{13}\\
1 & 0
\end{array}
\right\vert +\left\vert
\begin{array}
[c]{cc}%
j_{11} & j_{12}\\
j_{21} & j_{22}%
\end{array}
\right\vert =-j_{13}+j_{11}j_{22}-j_{12}j_{21}%
\]%
\[
c=-\det J=-\left\vert
\begin{array}
[c]{cc}%
j_{12} & j_{13}\\
j_{22} & j_{23}%
\end{array}
\right\vert =-j_{12}j_{23}+j_{13}j_{22}%
\]
such that the necessary and sufficient conditions for the local stability of a
given equilibrium are:%
\begin{align*}
1+a+b+c  &  >0\\
1-a+b-c  &  >0\\
1-b+ac-c^{2}  &  >0\\
3-b  &  >0
\end{align*}

At $P_{1}=\left(  -\Omega\Delta y^{BP},\Delta y^{BP}\right)  $:%
\begin{align*}
j_{11}  &  =1+\left(  \mu+\rho\right)  \left(  w^{C}+w^{E}\right) \\
j_{12}  &  =\left(  \mu+\rho\right)  w^{C}\Omega\\
j_{13}  &  =-\left(  \mu+\rho\right)  w^{E}\\
j_{21}  &  =-w^{flex}\beta\gamma\left(  w^{C}+w^{E}\right) \\
j_{22}  &  =1-w^{flex}\beta\left(  \gamma w^{C}\Omega+1\right) \\
j_{23}  &  =-w^{flex}\beta w^{E}\\
j_{31}  &  =1\\
j_{32}  &  =j_{33}=0
\end{align*}
so that%
\begin{align*}
a  &  =-j_{11}-j_{22}\\
&  =-1-\left(  \mu+\rho\right)  \left(  w^{C}+w^{E}\right)  -1+w^{flex}%
\beta\left(  \gamma w^{C}\Omega+1\right) \\
&  =-2-\left(  \mu+\rho\right)  \left(  w^{C}+w^{E}\right)  +w^{flex}%
\beta\left(  \gamma w^{C}\Omega+1\right)
\end{align*}%
\begin{align*}
b  &  =-j_{13}+j_{11}j_{22}-j_{12}j_{21}\\
&  =\left(  \mu+\rho\right)  w^{E}+\left[  1+\left(  \mu+\rho\right)  \left(
w^{C}+w^{E}\right)  \right]  \left[  1-w^{flex}\beta\left(  \gamma w^{C}%
\Omega+1\right)  \right] \\
&  +\left(  \mu+\rho\right)  w^{C}\Omega w^{flex}\beta\gamma\left(
w^{C}+w^{E}\right) \\
&  =\left(  \mu+\rho\right)  w^{E}+1+\left(  \mu+\rho\right)  \left(
w^{C}+w^{E}\right) \\
&  -w^{flex}\beta\left(  \gamma w^{C}\Omega+1\right)  -\left(  \mu
+\rho\right)  \left(  w^{C}+w^{E}\right)  w^{flex}\beta\left(  \gamma
w^{C}\Omega+1\right)
\end{align*}%
\begin{align*}
c  &  =-j_{12}j_{23}+j_{13}j_{22}\\
&  =\left(  \mu+\rho\right)  w^{C}\Omega w^{flex}\beta w^{E}-\left(  \mu
+\rho\right)  w^{E}\left[  1-w^{flex}\beta\left(  \gamma w^{C}\Omega+1\right)
\right] \\
&  =\left(  \mu+\rho\right)  w^{C}\Omega w^{flex}\beta w^{E}-\left(  \mu
+\rho\right)  w^{E}+\left(  \mu+\rho\right)  w^{E}w^{flex}\beta\left(  \gamma
w^{C}\Omega+1\right) \\
&  =\left(  \mu+\rho\right)  w^{C}\Omega w^{flex}\beta w^{E}-\left(  \mu
+\rho\right)  w^{E}\\
&  +\left(  \mu+\rho\right)  w^{E}w^{flex}\beta\gamma w^{C}\Omega+\left(
\mu+\rho\right)  w^{E}w^{flex}\beta\\
&  =\left(  \mu+\rho\right)  w^{E}\left(  w^{C}\Omega w^{flex}\beta
-1+w^{flex}\beta\gamma w^{C}\Omega+w^{flex}\beta\right)
\end{align*}
We can easily verify the first condition:
\begin{align*}
&  \text{(i) }1+a+b+c\\
&  =1-2-\left(  \mu+\rho\right)  \left(  w^{C}+w^{E}\right)  +w^{flex}%
\beta\left(  \gamma w^{C}\Omega+1\right) \\
&  +\left(  \mu+\rho\right)  w^{E}+1+\left(  \mu+\rho\right)  \left(
w^{C}+w^{E}\right) \\
&  -w^{flex}\beta\left(  \gamma w^{C}\Omega+1\right)  -\left(  \mu
+\rho\right)  \left(  w^{C}+w^{E}\right)  w^{flex}\beta\left(  \gamma
w^{C}\Omega+1\right) \\
&  +\left(  \mu+\rho\right)  w^{E}\left(  w^{C}\Omega w^{flex}\beta
-1+w^{flex}\beta\gamma w^{C}\Omega+w^{flex}\beta\right)
\end{align*}%
\begin{align*}
&  =\left(  \mu+\rho\right)  w^{E}-\left(  \mu+\rho\right)  \left(
w^{C}+w^{E}\right)  w^{flex}\beta\left(  \gamma w^{C}\Omega+1\right) \\
&  +\left(  \mu+\rho\right)  w^{E}\left(  w^{C}\Omega w^{flex}\beta
-1+w^{flex}\beta\gamma w^{C}\Omega+w^{flex}\beta\right)
\end{align*}%
\begin{align*}
&  =-\left(  \mu+\rho\right)  \left(  w^{C}+w^{E}\right)  w^{flex}\beta\gamma
w^{C}\Omega\\
&  -\left(  \mu+\rho\right)  \left(  w^{C}+w^{E}\right)  w^{flex}\beta\\
&  +\left(  \mu+\rho\right)  w^{E}w^{C}\Omega w^{flex}\beta+\left(  \mu
+\rho\right)  w^{E}w^{flex}\beta\gamma w^{C}\Omega+\left(  \mu+\rho\right)
w^{E}w^{flex}\beta
\end{align*}%
\begin{align*}
&  =-\left(  \mu+\rho\right)  \left(  w^{C}+w^{E}\right)  w^{flex}\beta\gamma
w^{C}\Omega\\
&  -\left(  \mu+\rho\right)  w^{C}w^{flex}\beta-\left(  \mu+\rho\right)
w^{E}w^{flex}\beta\\
&  +\left(  \mu+\rho\right)  w^{E}w^{C}\Omega w^{flex}\beta+\left(  \mu
+\rho\right)  w^{E}w^{flex}\beta\gamma w^{C}\Omega+\left(  \mu+\rho\right)
w^{E}w^{flex}\beta
\end{align*}%
\begin{align*}
&  =-\left(  \mu+\rho\right)  \left(  w^{C}+w^{E}\right)  w^{flex}\beta\gamma
w^{C}\Omega\\
&  -\left(  \mu+\rho\right)  w^{C}w^{flex}\beta+\left(  \mu+\rho\right)
w^{E}w^{C}\Omega w^{flex}\beta+\left(  \mu+\rho\right)  w^{E}w^{flex}%
\beta\gamma w^{C}\Omega
\end{align*}%
\begin{align*}
&  =-\left(  \mu+\rho\right)  w^{C}w^{flex}\beta\gamma w^{C}\Omega-\left(
\mu+\rho\right)  w^{E}w^{flex}\beta\gamma w^{C}\Omega\\
&  -\left(  \mu+\rho\right)  w^{C}w^{flex}\beta+\left(  \mu+\rho\right)
w^{E}w^{C}\Omega w^{flex}\beta+\left(  \mu+\rho\right)  w^{E}w^{flex}%
\beta\gamma w^{C}\Omega
\end{align*}%
\begin{align*}
&  =-\left(  \mu+\rho\right)  w^{C}w^{flex}\beta\gamma w^{C}\Omega\\
&  -\left(  \mu+\rho\right)  w^{C}w^{flex}\beta+\left(  \mu+\rho\right)
w^{C}w^{flex}\beta w^{E}\Omega
\end{align*}%
\[
=-\left(  \mu+\rho\right)  w^{C}w^{flex}\beta\left(  \gamma w^{C}%
\Omega+1-w^{E}\Omega\right)<0
\]
is always violated. Therefore, $P_1$ is unstable.

At $P_{2,3}=\left(  \bar{e}_{2,3},\Delta y^{BP}\right)  =\left(  -\Omega\Delta
y^{BP}\pm\sqrt{\frac{w^{C}}{w^{F}}},\Delta y^{BP}\right)  $, noticing that%
\[
\left(  -\bar{e}_{2,3}-\Omega\Delta y^{BP}\right)  ^{2}=\left(  \Omega\Delta
y^{BP}\pm\sqrt{\frac{w^{C}}{w^{F}}}-\Omega\Delta y^{BP}\right)  ^{2}%
=\frac{w^{C}}{w^{F}}%
\]
the elements of the Jacobian matrix become:%
\begin{align*}
j_{11}  &  =1+\left(  \mu+\rho\right)  \left(  w^{E}-2w^{C}\right) \\
j_{12}  &  =-2\left(  \mu+\rho\right)  \Omega w^{C}\\
j_{13}  &  =-\left(  \mu+\rho\right)  w^{E}\\
j_{21}  &  =-w^{flex}\beta\gamma\left(  w^{E}-2w^{C}\right) \\
j_{22}  &  =1-w^{flex}\beta\left(  1-2\gamma\Omega w^{C}\right) \\
j_{23}  &  =-w^{flex}\beta w^{E}\\
j_{31}  &  =1\\
j_{32}  &  =j_{33}=0
\end{align*}

Therefore%
\begin{align*}
a &  =-j_{11}-j_{22}\\
&  =-1-\left(  \mu+\rho\right)  \left(  w^{E}-2w^{C}\right)  -1+w^{flex}%
\beta\left(  1-2\gamma\Omega w^{C}\right)  \\
&  =-2-\left(  \mu+\rho\right)  \left(  w^{E}-2w^{C}\right)  +w^{flex}%
\beta\left(  1-2\gamma\Omega w^{C}\right)
\end{align*}%
\begin{align*}
b &  =-j_{13}+j_{11}j_{22}-j_{12}j_{21}\\
&  =\left(  \mu+\rho\right)  w^{E}+\left[  1+\left(  \mu+\rho\right)  \left(
w^{E}-2w^{C}\right)  \right]  \\
&  \times\left[  1+2w^{flex}\beta\gamma\Omega w^{C}-w^{flex}\beta\right]
-2\left(  \mu+\rho\right)  \Omega w^{C}w^{flex}\beta\gamma\left(  w^{E}%
-2w^{C}\right)  \\
&  =\left(  \mu+\rho\right)  w^{E}+1+2w^{flex}\beta\gamma\Omega w^{C}%
-w^{flex}\beta+\left(  \mu+\rho\right)  w^{E}\left(  1-w^{flex}\beta\right)
\\
&  -2\left(  \mu+\rho\right)  w^{C}\left(  1-w^{flex}\beta\right)  \\
&  =2\left(  \mu+\rho\right)  w^{E}+1+2w^{flex}\beta\gamma\Omega
w^{C}-w^{flex}\beta-\left(  \mu+\rho\right)  w^{E}w^{flex}\beta-2\left(
\mu+\rho\right)  w^{C}\left(  1-w^{flex}\beta\right)
\end{align*}%
\begin{align*}
c &  =-j_{12}j_{23}+j_{13}j_{22}\\
&  =-2\left(  \mu+\rho\right)  \Omega w^{C}w^{flex}\beta w^{E}-\left(
\mu+\rho\right)  w^{E}\left[  1+2w^{flex}\beta\gamma\Omega w^{C}-w^{flex}%
\beta\right]  \\
&  =-2\left(  \mu+\rho\right)  \Omega w^{C}w^{flex}\beta w^{E}\left(
1+\gamma\right)  -\left(  \mu+\rho\right)  w^{E}\left(  1-w^{flex}%
\beta\right)  <0
\end{align*}

\bigskip

Then, the local stability conditions are given by:%
\begin{align*}
&  \text{(i) }1+a+b+c\\
&  =1-2-\left(  \mu+\rho\right)  \left(  w^{E}-2w^{C}\right)  -2w^{flex}%
\beta\gamma\Omega w^{C}+w^{flex}\beta\\
&  +2\left(  \mu+\rho\right)  w^{E}+1+2w^{flex}\beta\gamma\Omega
w^{C}-w^{flex}\beta\\
&  -\left(  \mu+\rho\right)  w^{E}w^{flex}\beta-2\left(  \mu+\rho\right)
w^{C}\left(  1-w^{flex}\beta\right)  \\
&  -2\left(  \mu+\rho\right)  \Omega w^{C}w^{flex}\beta w^{E}\left(
1+\gamma\right)  -\left(  \mu+\rho\right)  w^{E}\left(  1-w^{flex}%
\beta\right)  \\
&  =-\left(  \mu+\rho\right)  \left(  w^{E}-2w^{C}\right)  +2\left(  \mu
+\rho\right)  w^{E}-2\left(  \mu+\rho\right)  w^{C}\left(  1-w^{flex}%
\beta\right)  \\
&  -2\left(  \mu+\rho\right)  \Omega w^{C}w^{flex}\beta w^{E}\left(
1+\gamma\right)  -\left(  \mu+\rho\right)  w^{E}\\
&  =2\left(  \mu+\rho\right)  w^{C}w^{flex}\beta\left[  1-\Omega w^{E}\left(
1+\gamma\right)  \right]  \gtreqqless0
\end{align*}%
\begin{align*}
&  \text{(ii) }1-a+b-c\\
&  =1+2+\left(  \mu+\rho\right)  \left(  w^{E}-2w^{C}\right)  +2w^{flex}%
\beta\gamma\Omega w^{C}-w^{flex}\beta\\
&  +2\left(  \mu+\rho\right)  w^{E}+1+2w^{flex}\beta\gamma\Omega
w^{C}-w^{flex}\beta-\left(  \mu+\rho\right)  w^{E}w^{flex}\beta-2\left(
\mu+\rho\right)  w^{C}\left(  1-w^{flex}\beta\right)  \\
&  +2\left(  \mu+\rho\right)  \Omega w^{C}w^{flex}\beta w^{E}\left(
1+\gamma\right)  +\left(  \mu+\rho\right)  w^{E}\left(  1-w^{flex}%
\beta\right)  \\
&  =4-4\left(  \mu+\rho\right)  w^{C}-2w^{flex}\beta\left(  1-2\gamma\Omega
w^{C}\right)  +2\left(  \mu+\rho\right)  w^{C}w^{flex}\beta\\
&  +4\left(  \mu+\rho\right)  w^{E}+2\left(  \mu+\rho\right)  \Omega
w^{C}w^{flex}\beta w^{E}\left(  1+\gamma\right)  -2\left(  \mu+\rho\right)
w^{E}w^{flex}\beta\\
&  =4-4\left(  \mu+\rho\right)  w^{C}-2w^{flex}\beta\left(  1-2\gamma\Omega
w^{C}\right)  +2\left(  \mu+\rho\right)  w^{C}w^{flex}\beta\\
&  +2\left(  \mu+\rho\right)  \left[  2+\Omega w^{C}w^{flex}\beta\left(
1+\gamma\right)  -w^{flex}\beta\right]  w^{E}\\
&  =A+\underset{>0}{\underbrace{2\left(  \mu+\rho\right)  \left[  2+\Omega
w^{C}w^{flex}\beta\left(  1+\gamma\right)  -w^{flex}\beta\right]  w^{E}}}%
\end{align*}%
\begin{align*}
&  \text{(iii) }1-b+ac-c^{2}\\
&  =1-2\left(  \mu+\rho\right)  w^{E}-1-2w^{flex}\beta\gamma\Omega
w^{C}+w^{flex}\beta+\left(  \mu+\rho\right)  w^{E}w^{flex}\beta+2\left(
\mu+\rho\right)  w^{C}\left(  1-w^{flex}\beta\right)  \\
&  +\left[  -2-\left(  \mu+\rho\right)  \left(  w^{E}-2w^{C}\right)
-2w^{flex}\beta\gamma\Omega w^{C}+w^{flex}\beta\right]  \\
&  \times\left[  -2\left(  \mu+\rho\right)  \Omega w^{C}w^{flex}\beta
w^{E}\left(  1+\gamma\right)  -\left(  \mu+\rho\right)  w^{E}\left(
1-w^{flex}\beta\right)  \right]  \\
&  -\left[  -2\left(  \mu+\rho\right)  \Omega w^{C}w^{flex}\beta w^{E}\left(
1+\gamma\right)  -\left(  \mu+\rho\right)  w^{E}\left(  1-w^{flex}%
\beta\right)  \right]  ^{2}\\
&  =-2\left(  \mu+\rho\right)  w^{E}-2w^{flex}\beta\gamma\Omega w^{C}%
+w^{flex}\beta+\left(  \mu+\rho\right)  w^{E}w^{flex}\beta+2\left(  \mu
+\rho\right)  w^{C}\left(  1-w^{flex}\beta\right)  \\
&  -2\left[  -2\left(  \mu+\rho\right)  \Omega w^{C}w^{flex}\beta w^{E}\left(
1+\gamma\right)  -\left(  \mu+\rho\right)  w^{E}\left(  1-w^{flex}%
\beta\right)  \right]  \\
&  -\left(  \mu+\rho\right)  \left(  w^{E}-2w^{C}\right)  \left[  -2\left(
\mu+\rho\right)  \Omega w^{C}w^{flex}\beta w^{E}\left(  1+\gamma\right)
-\left(  \mu+\rho\right)  w^{E}\left(  1-w^{flex}\beta\right)  \right]  \\
&  -2w^{flex}\beta\gamma\Omega w^{C}\left[  -2\left(  \mu+\rho\right)  \Omega
w^{C}w^{flex}\beta w^{E}\left(  1+\gamma\right)  -\left(  \mu+\rho\right)
w^{E}\left(  1-w^{flex}\beta\right)  \right]  \\
&  +w^{flex}\beta\left[  -2\left(  \mu+\rho\right)  \Omega w^{C}w^{flex}\beta
w^{E}\left(  1+\gamma\right)  -\left(  \mu+\rho\right)  w^{E}\left(
1-w^{flex}\beta\right)  \right]  \\
&  -\left[  2\left(  \mu+\rho\right)  \Omega w^{C}w^{flex}\beta w^{E}\left(
1+\gamma\right)  \right]  ^{2}\\
&  -4\left(  \mu+\rho\right)  \Omega w^{C}w^{flex}\beta w^{E}\left(
1+\gamma\right)  \left(  \mu+\rho\right)  w^{E}\left(  1-w^{flex}\beta\right)
\\
&  -\left[  \left(  \mu+\rho\right)  w^{E}\left(  1-w^{flex}\beta\right)
\right]  ^{2}%
\end{align*}%

After many tedious algebraic manipulations, condition (iii) can be rewritten as:
\begin{gather*}
=B+4\left(  \mu+\rho\right)  \Omega w^{C}w^{flex}\beta\left(  1+\gamma\right)
w^{E}-\left(  \mu+\rho\right)  w^{flex}\beta w^{E}\\
-4\left(  \mu+\rho\right)  ^{2}\Omega w^{C^{2}}w^{flex}\beta\left(
1+\gamma\right)  w^{E}-2\left(  \mu+\rho\right)  ^{2}\left(  1-w^{flex}%
\beta\right)  w^{C}w^{E}\\
+4w^{flex}\beta\gamma\Omega w^{C}\left(  \mu+\rho\right)  \Omega w^{C}%
w^{flex}\beta\left(  1+\gamma\right)  w^{E}\\
+2w^{flex}\beta\gamma\Omega w^{C}\left(  \mu+\rho\right)  \left(
1-w^{flex}\beta\right)  w^{E}\\
-2\left(  \mu+\rho\right)  \Omega w^{C}w^{flex}\beta\left(  1+\gamma\right)
w^{flex}\beta w^{E}\\
-\left(  \mu+\rho\right)  \left(  1-w^{flex}\beta\right)  w^{flex}\beta
w^{E}\\
+\left(  \mu+\rho\right)  \left(  \mu+\rho\right)  w^{flex}\beta\left[
2\Omega w^{C}w^{flex}\beta\left(  1+\gamma\right)  +\left(  1-w^{flex}%
\beta\right)  \right]  \left[  1-2\Omega w^{C}\left(  1+\gamma\right)
\right]  w^{E^{2}}%
\end{gather*}

Notice that when $w^E=0$, conditions (i)-(iii) are identical to those of system (\ref{2D}). It is also clear from (ii) that extrapolators make the Flip bifurcation more difficult to occur. Finally, condition (iii) is a quadratic function in $w^E$ and will be U-shaped provided that:
\begin{equation*}
    2\Omega w^{C}\left(  1+\gamma\right)<1
\end{equation*}
or inverted U-shaped when the opposite inequality holds. A violation of (i) while (ii)-(iii) are satisfied is associated with a Fold bifurcation. When (ii) is violated while (i) and (iii) are satisfied, a Flip bifurcation occurs. Finally, when only (iii) is violated, a Neimark-Sacker bifurcation may happen (see \href{#Lines et al 2020}{Lines et al., 2020}). Given our choice of parameters, only a Flip bifurcation can and is shown to occur.

\begin{figure}[tbp]
    \centering
    \includegraphics[width=0.5\linewidth]{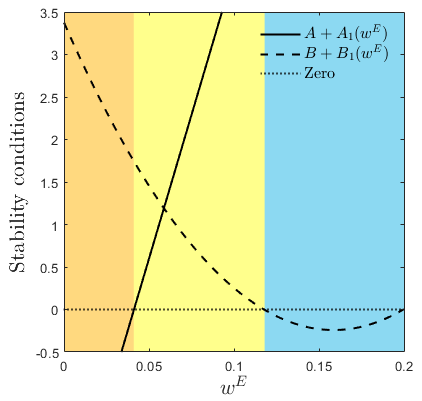}
    \caption{Flip and Neimark-Sacker Bifurcations. Parameters $\Omega=0.01$, $\theta=0.3$, $\pi=2$, $\Delta y^{BP}$, $w^F=0.8$, $w^C=1-w^F-w^E$, $\mu=4.5$, $\rho=4$, $w^{flex}=0.1$, $\beta=0.1$. Dark yellow indicates values of $w^E$ for which $A+A_1<0$ and $B+B_1>0$, so that a Flip bifurcation occurred. Light yellow marks the region in the parameter space for which $A+A_1>0$ and $B+B_1>0$. Finally, in blue, we have those for which $A+A_1 > 0$ and $B+B_1 < 0$, so that a Neimark-Sacker bifurcation may occur.}
    \label{Bifurcations}
\end{figure}

\end{document}